\documentclass[%
reprint,
superscriptaddress,
 amsmath,amssymb,
 aps,
prb,
]{revtex4-2}

\usepackage[table]{xcolor}

\usepackage{graphicx            }
\usepackage{dcolumn}            
\usepackage{bm}                 
\usepackage[utf8x]{inputenc}    
\usepackage{physics}
\usepackage{xcolor,soul}
\usepackage[colorlinks,citecolor=red,urlcolor=blue,bookmarks=false,hypertexnames=true]{hyperref} 
\usepackage{subfig}
\usepackage{tabularx}
\usepackage{multirow}
\usepackage{booktabs}
\usepackage{enumitem}           
\usepackage{siunitx}            

\begin{document}
\preprint{APS/123-QED}
\title{Optical and excitonic properties of transition metal oxide perovskites\\by the Bethe-Salpeter equation}

\author{Lorenzo Varrassi}
\affiliation{Dipartimento di Fisica e Astronomia, Università di Bologna, 40127 Bologna, Italy\looseness=-1}
\author{Peitao Liu}
\affiliation{University of Vienna, Faculty of Physics and Center for Computational Materials Science, Kolingasse 14-16, A-1090, Vienna, Austria}%
\author{Zeynep Erg{\"o}nenc Yavas}
\affiliation{Turkish Aerospace Industries Inc. - Department of Materials Engineering -  Fethiye Mahallesi, Havacılık Bulvarı No:17 06980 Kazan-Ankara, Turkey}
\author{Menno Bokdam}
\affiliation{University of Twente, Faculty of Science and Technology
and MESA+ Institute
Enschede, the Netherlands} %
\author{Georg Kresse}
\affiliation{University of Vienna, Faculty of Physics and Center for Computational Materials Science, Kolingasse 14-16, A-1090, Vienna, Austria}%
\author{Cesare Franchini}
\affiliation{Dipartimento di Fisica e Astronomia, Università di Bologna, 40127 Bologna, Italy\looseness=-1}
\affiliation{University of Vienna, Faculty of Physics and Center for Computational Materials Science, Kolingasse 14-16, A-1090, Vienna, Austria}%
\date{\today}

\begin{abstract}
We present a systematic investigation of the role and importance of excitonic effects on the optical properties of transitions metal oxide perovskites. 
A representative set of fourteen compounds has been selected, including 3$d$ (SrTiO$_3$, LaScO$_3$, LaTiO$_3$, LaVO$_3$, LaCrO$_3$, LaMnO$_3$, LaFeO$_3$ and SrMnO$_3$), 4$d$ (SrZrO$_3$, SrTcO$_3$ and Ca$_2$RuO$_4$) and 5$d$ (SrHfO$_3$, KTaO$_3$ and NaOsO$_3$) perovskites, covering a band gap ranging from 0.1 eV to 6.1 eV and exhibiting different electronic, structural and magnetic properties.
Optical conductivities and optical transitions including electron-hole interactions are calculated 
through the solution of the Bethe-Salpeter equation (BSE) with quasi-particle energies evaluated by single-shot $G_0W_0$ approximation. 
The exciton binding energies are computed by means of a model-BSE (mBSE), carefully benchmarked against the full BSE method, in order to obtain well-converged results in terms of k-point sampling. 
The predicted results are compared with available measured data, with an overall satisfactory agreement between theory and experiment.
\end{abstract}

\maketitle
\section{\label{sec:Introduction}Introduction}

\begin{table*}[!ht]
\centering
\begin{tabularx}{\textwidth}{X|XXX|XXXXX}
\toprule\toprule
                                   &\quad \begin{tabular}[c]{@{}c@{}}Crystal\\ Structures\end{tabular} & \begin{tabular}[c]{@{}c@{}}Electronic\\ Configurations\end{tabular} & \hspace{0.3cm}\begin{tabular}[c]{@{}c@{}}Magnetic\\ Orderings\end{tabular} &   \hspace{0.45cm}$E_{pw}$(eV) & 
                                   \hspace{0.3cm}$N_{pw}$ & \hspace{0.3cm}$N_\omega$ & $N_O$ & $N_C$ \\ \midrule
\multicolumn{1}{c|}{\hspace{0.3cm}SrTiO$_3$\hspace{0.4cm}}     &\hspace{0.4cm} C-P$_{m \bar{3} m}$    & \hspace{0.65cm}${3}d^0$                & \multicolumn{1}{c|}{NM}       & \hspace{0.6cm}600      & \hspace{0.3cm}512      & \hspace{0.3cm}96         & 12    & 10    \\
\multicolumn{1}{c|}{\hspace{0.3cm}SrZrO$_3$\hspace{0.3cm}}     &\hspace{0.4cm} C-P$_{m \bar{3} m}$    & \hspace{0.65cm}{4}$d^0$                & \multicolumn{1}{c|}{NM}       & \hspace{0.6cm}650      & \hspace{0.3cm}1972     & \hspace{0.3cm}64         & 12    & 12    \\
\multicolumn{1}{c|}{\hspace{0.3cm}SrHfO$_3$\hspace{0.3cm}}     &\hspace{0.4cm} C-P$_{m \bar{3} m}$    & \hspace{0.65cm}{5}$d^0$                & \multicolumn{1}{c|}{NM}       & \hspace{0.6cm}650      & \hspace{0.3cm}2304     & \hspace{0.3cm}96         & 12    & 13    \\
\multicolumn{1}{c|}{\hspace{0.3cm}KTaO$_3$\hspace{0.3cm}}      &\hspace{0.4cm} C-P$_{m \bar{3} m}$    & \hspace{0.65cm}{5}$d^0$                & \multicolumn{1}{c|}{NM}       & \hspace{0.6cm}500      & \hspace{0.3cm}896      & \hspace{0.3cm}96         & 12    & 12    \\
\multicolumn{1}{c|}{\hspace{0.3cm}LaScO$_3$\hspace{0.3cm}}     &\hspace{0.4cm} O-P$_{nma}$            & \hspace{0.65cm}{3}$d^0$                & \multicolumn{1}{c|}{NM}       & \hspace{0.6cm}500      & \hspace{0.3cm}1280     & \hspace{0.3cm}64         & 32    & 32    \\
\multicolumn{1}{c|}{\hspace{0.3cm}LaTiO$_3$\hspace{0.3cm}}     &\hspace{0.4cm} O-P$_{nma}$            & \hspace{0.65cm}{3}$t_{2g}^1$           & \multicolumn{1}{c|}{G-AFM}    & \hspace{0.6cm}500      & \hspace{0.3cm}448      & \hspace{0.3cm}64         & 34    & 34    \\
\multicolumn{1}{c|}{\hspace{0.3cm}LaVO$_3$\hspace{0.3cm}}      &\hspace{0.4cm} M-P$_{21/b}$           & \hspace{0.65cm}{3}$t_{2g}^2$           & \multicolumn{1}{c|}{G-AFM}    & \hspace{0.6cm}500      & \hspace{0.3cm}448      & \hspace{0.3cm}64         & 30    & 30    \\
\multicolumn{1}{c|}{\hspace{0.3cm}LaCrO$_3$\hspace{0.3cm}}     &\hspace{0.4cm} O-P$_{nma}$            & \hspace{0.65cm}{3}$t_{2g}^3$           & \multicolumn{1}{c|}{G-AFM}    & \hspace{0.6cm}500      & \hspace{0.3cm}448      & \hspace{0.3cm}64         & 32    & 32    \\
\multicolumn{1}{c|}{\hspace{0.3cm}LaMnO$_3$\hspace{0.3cm}}     &\hspace{0.4cm} O-P$_{nma}$            & \hspace{0.65cm}{3}$t_{2g}^3e_g^1$      & \multicolumn{1}{c|}{A-AFM}    & \hspace{0.6cm}500      &\hspace{0.3cm}448       & \hspace{0.3cm}64         & 26    & 26    \\
\multicolumn{1}{c|}{\hspace{0.3cm}LaFeO$_3$\hspace{0.3cm}}     &\hspace{0.4cm} O-P$_{nma}$            & \hspace{0.65cm}{3}$t_{2g}^3e_g^2$      & \multicolumn{1}{c|}{G-AFM}    & \hspace{0.6cm}500      & \hspace{0.3cm}448      & \hspace{0.3cm}96         & 34    & 34    \\
\multicolumn{1}{c|}{\hspace{0.3cm}SrMnO$_3$\hspace{0.3cm}}     &\hspace{0.4cm} C-P$_{m \bar{3} m}$    & \hspace{0.65cm}{3}$t_{2g}^3$           & \multicolumn{1}{c|}{G-AFM}    & \hspace{0.6cm}500      & \hspace{0.3cm}448      & \hspace{0.3cm}64         & 29    & 29    \\
\multicolumn{1}{c|}{\hspace{0.3cm}SrTcO$_3$\hspace{0.3cm}}     &\hspace{0.4cm} O-P$_{nma}$            & \hspace{0.65cm}{4}$t_{2g}^3$           & \multicolumn{1}{c|}{G-AFM}    & \hspace{0.6cm}500      & \hspace{0.3cm}512      & \hspace{0.3cm}64         & 30    & 30    \\
\multicolumn{1}{c|}{\hspace{0.3cm}Ca$_2$RuO$_4$\hspace{0.3cm}} &\hspace{0.4cm} O-P$_{nma}$            & \hspace{0.65cm}{4}$t_{2g}^3e_g^1$      & \multicolumn{1}{c|}{AFM}      & \hspace{0.6cm}500      & \hspace{0.3cm}512      & \hspace{0.3cm}64         & 30    & 37    \\
\multicolumn{1}{c|}{\hspace{0.3cm}NaOsO$_3$\hspace{0.3cm}}     &\hspace{0.4cm} O-P$_{bca}$            & \hspace{0.65cm}{5}$t_{2g}^3$           & \multicolumn{1}{c|}{G-AFM}    & \hspace{0.6cm}500      & \hspace{0.3cm}448      & \hspace{0.3cm}64         & 30    & 30    \\ \bottomrule\bottomrule
\end{tabularx}
\caption{Material dataset and main computational parameters.
The first column lists the considered compounds.
The second set of columns collects informations on the crystal structures (C=cubic, T=tetragonal, O=orthorombic, M=monoclinic), electronic configurations of the transition metal $d$ shell and ground state magnetic orderings (NM=non-magnetic and different types of anti-ferromagnetic spin configurations~\cite{He.PRB.86.235117}).
The last set of columns lists the relevant computational parameters: plane-wave energy cutoff for orbitals ($E_{pw}$), number of bands ($N_{pw}$), number of frequency points used for the calculation of the $GW$ polarizability ($N_{\omega}$ ). $N_O$ and $N_C$ refer to the number of occupied and conduction bands, respectively, included in the BSE equation. The plane-wave energy cutoff for the response function is chosen to be $2/3E_{pw}$.}
\label{tab:CompDetails:Param}
\end{table*}
The study of transition metal oxide (TMO) perovskites has brought to light a wide array of physical and chemical properties, including colossal magnetoresistance  ~\cite{PhysRevLett.71.2331,RevModPhys.73.583}
, multiferroicity~\cite{doi:10.1080}, metal-insulator transitions~\cite{RevModPhys.70.1039}, superconductivity~\cite{Bednorz1986,doi:10.1063}, two dimensional electron gas~\cite{Ohtomo2004} and spin and charge ordering~\cite{Rao2000}. In the last decades the 4$d$ and 5$d$ TMO perovskites have gained increasing interest due to the discovery of novel electronic and magnetic quantum states of matter arising from the coupling between spin-orbit interaction and other active degrees of freedom~\cite{doi:10.1146,Martins,PhysRevMaterials.4.045001,app11062527}.
\newline Up to now, few theoretical studies have investigated the role of excitonic effects on the optical spectra of TMO perovskites~\cite{PhysRevB.89.045104,Sponza.PRB.87.235102,Liu.PRM.2.075003,Begum.PRM.3.065004,PhysRevB.101.155135}. These works have proved that the random phase approximation (RPA) is able to reproduce the experimental data only to a limited extent and that  the inclusion of electron-hole (e-h) interaction is often pivotal to achieve a satisfying account of the optical transitions~\cite{Sponza.PRB.87.235102,Liu.PRM.2.075003,Begum.PRM.3.065004}.
This paper attempts to extend the study of excitonic effects from individual compounds towards a larger representative dataset, aiming to contribute to a comprehensive understanding of the role the electron-hole interaction in TMO perovskites.

In order to compute the optical properties, we  solve the Bethe-Salpeter equation (BSE)~\cite{RevModPhys.74.601,BSE.PhysRevLett.80.4510} through a direct diagonalization scheme, which offers direct access to excitonic wavefunctions and allows for a more transparent interpretation of the main features of the optical spectra. To correctly estimate the interband optical transition energies it is necessary to accurately calculate the quasi-particle energies and, above all, the fundamental gaps. 
Density functional theory (DFT) is usually considered not suitable in this regard; it commonly underestimates fundamental gaps and does not provide a reliable account of the excited state properties~\cite{PhysRevLett.96.226402}. This task is instead successfully achieved by the $GW$ approximation~\cite{PhysRev.139.A796,PhysRevB.25.2867,PhysRevLett.55.1418}, which provides a good description of insulating gaps and band dispersions for the TMO perovskites~\cite{Ergorenc.PRM.2.024601,Nohara.PRB.79.195110,PhysRevB.87.085112,Franchini_2012,PhysRevB.93.075125}. To accurately predict optical properties it is therefore natural to combine $GW$ and BSE in a single computational protocol. 
\newline However, the resulting $GW$+BSE procedure is computationally very demanding. 
On the one hand, both $GW$ and BSE are notoriously computationally expensive, with an unfavourable scaling (both standard implementations exhibit at least quartic scaling in the system size and quadratic in the number of k-points~\cite{PhysRevB.59.5441}): their application to large systems therefore presents a technical challenge. On the other hand, to obtain reliable results, a precise convergence procedure is needed  as pointed out in seminal BSE studies~\cite{PhysRevLett.80.4510, PhysRevLett.83.3971}.
For example, a too sparse k-point mesh may introduce spurious artefacts~\cite{Laskowski.PRB.72.035204,PhysRevB.84.085145} or incorrect estimations of the exciton binding energy~\cite{Fuchs.PRB.78.085103,Bokdam.SR.2016.28618}. 
To bypass these limitations alternative schemes have been proposed, from shifted~\cite{PhysRevB.59.5441,Kammerlander.PRB.86.125203} or hybrid k-point meshes~\cite{Fuchs.PRB.78.085103} to interpolation schemes~\cite{PhysRevB.62.4927} or methods based on density matrix perturbation theory~\cite{PhysRevB.85.045116}.
\newline Additional complications arise from the fact that the considered materials cover a wide range of electronic and excitonic behaviours, with gaps ranging from 0.1 eV to 6.0 eV and originating from different mechanisms (band insulator, Mott-Hubbard, Mott-Dirac, etc.). This unavoidably affects the nature and properties of the exciton wavefunctions.
These fundamental characteristics have in turn a strong impact on the convergence rate of optical properties~\cite{Laskowski.PRB.72.035204}, especially with respect to the k-point sampling of the Brillouin zone (BZ).  Therefore, in order to achieve converged and reliable exciton binding energies and ensure equally well-defined results for all compounds in the dataset, we adopted a model-BSE (mBSE) procedure, which uses a parametrized model for the dielectric screening~\cite{Bokdam.SR.2016.28618, Liu.PRM.2.075003}.

Following the work of  He \textit{et al.}~\cite{He.PRB.86.235117} and Erg{\"o}renc \textit{et al.}~\cite{Ergorenc.PRM.2.024601}, 
we adopt a representative dataset of fourteen TM perovskites that includes compounds with different electronic, structural, magnetic and dielectric properties, as summarized in Table~\ref{tab:CompDetails:Param} and in Refs.~\cite{Ergorenc.PRM.2.024601}. Specifically: (i) insulating gaps ranging from 0.1 eV to 6 eV; (ii) 3$d$, 4$d$ and 5$d$ TM-based perovskites with different orbital occupancy; (iii) non-magnetic and differently ordered AFM patterns; (iv) various crystal structures with different types of internal structural distortions (e.g. with and without Jahn-Teller instabilities); (v) macroscopic dielectric constant from 1 to 10.
The complete atomic positions, as well as the experimental sources are given in the Supplementary Materials (SM)~\cite{SupplementaryMaterial}.
We note that in order to correctly describe the G-AFM magnetic ordering a supercell containing four formula units was used.

The paper is organized as follows: Section~\ref{sec:CompDetails} describes the computational setup, with a particular focus on the k-point convergence procedure. The discussion of the optical properties is divided in three sections, each for a specific subgroup of materials: Section~\ref{sec:CubicPerov} for the $C$-$P_{m \bar{3} m}$ perovskites, \ref{sec:LaPerov} for the LaTMO$_3$ series and \ref{sec:RemainingPerov} for Ca$_2$RuO$_4$, NaOsO$_3$, SrMnO$_3$ and SrTcO$_3$, followed by the conclusion in Section~\ref{sec:conclude}.
\begin{figure*}[!hbt]
	\includegraphics[width=17.6cm]{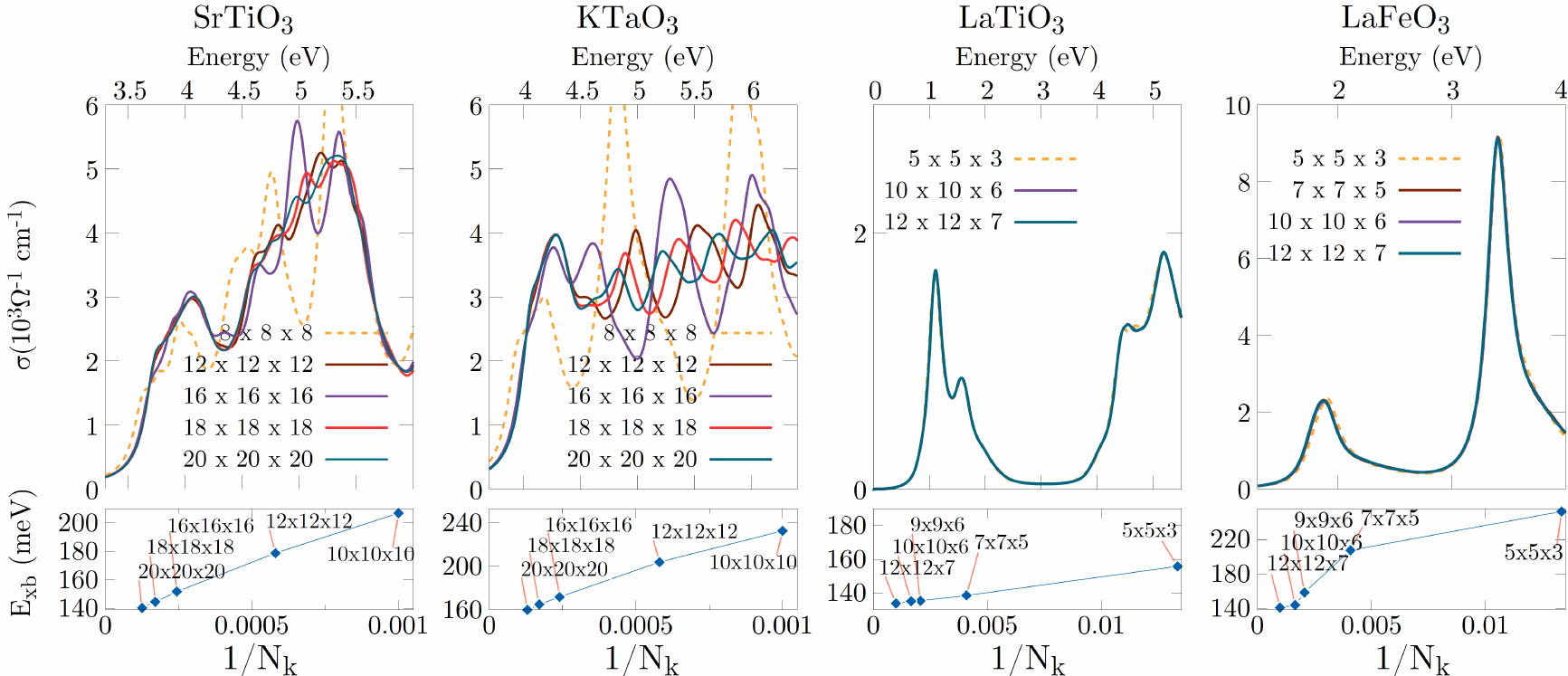} 
	\caption{\label{fig:CompDetails:KptsConvergencExample}
	Convergence tests for the mBSE derived $\sigma(\omega)$ (top panels) and exciton binding energies $E_{xb}$ (lower panels) with respect to the number of k-points. The optical conductivities are expressed in $10^{-3} \Omega^{-1}$cm$^{-1}$.
	}
\end{figure*}

\section{\label{sec:CompDetails}Computational Details}
All \textit{ab initio} calculations were performed using the Vienna \textit{ab initio} Simulation Package (VASP)~\cite{PhysRevB.47.558,PhysRevB.54.11169} with the augmented wave method (PAW)~\cite{PhysRevB.50.17953}. The potential types are listed in Table~\ref{tab:CompDetails:POTCAR}; the GW versions of all PAW potentials were used. The ultrasoft (US) versions of the potentials were used for all materials except for Sr and Ti in SrTiO$_3$, for which the norm-conserving (NC) versions were used, consistently with Erg{\"o}renc \textit{et al.}~\cite{Ergorenc.PRM.2.024601}.
\begin{table}[h]
\centering
\begin{tabularx}{\columnwidth}{@{}>{\centering\arraybackslash}>{\hsize=1.2cm}X >{\hsize=1.6cm}X XXXXX@{}}
\toprule\toprule
Element & PAW & $r_s$ & $r_p$ & $r_d$ & $r_f$ & $E_{pw}$ \\ \midrule
O       & GW-US  & 1.2   & 1.5   & 1.6   & 1.4   & 434      \\
        & GW-NC  & 1.0   & 1.1   & 1.1   &       & 765      \\
Na      & GW-US  & 1.6   & 2.0   & 2.2   &       & 260      \\
K       & GW-US  & 1.7   & 2.0   & 2.5   &       & 249      \\
Ca      & GW-US  & 1.6   & 1.9   & 2.2   &       & 281      \\
Sc      & GW-US  & 1.7   & 1.7   & 1.9   & 2.0   & 379      \\
Ti      & GW-US  & 1.7   & 1.7   & 2.0   & 2.0   & 384      \\
        & GW-NC  & 0.9   & 1.4   & 1.9   & 1.9   & 785      \\
V       & GW-US  & 1.8   & 1.7   & 1.9   & 2.0   & 382      \\
Cr      & GW-US  & 2.8   & 2.5   & 2.5   & 2.8   & 219      \\
Mn      & GW-US  & 1.6   & 1.7   & 1.9   & 1.9   & 385      \\
Fe      & GW-US  & 1.5   & 1.7   & 1.9   & 2.0   & 388      \\
Sr      & GW-US  & 1.7   & 2.1   & 2.5   & 2.5   & 225      \\
        & GW-NC  & 1.1   & 2.0   & 2.3   & 2.1   & 543      \\
Zr      & GW-US  & 1.3   & 1.8   & 2.0   & 2.1   & 346      \\
Tc      & GW-US  & 1.5   & 1.8   & 2.2   & 2.3   & 318      \\
Ru      & GW-US  & 1.5   & 1.8   & 2.2   & 2.3   & 321      \\
La      & GW-US  & 1.6   & 1.8   & 2.2   & 2.5   & 314      \\
Hf      & GW-US  & 1.5   & 1.9   & 2.2   & 2.5   & 283      \\
Ta      & GW-US  & 1.5   & 1.9   & 2.2   & 2.5   & 286      \\
Os      & GW-US  & 1.5   & 1.8   & 2.2   & 2.3   & 319      \\ \bottomrule\bottomrule
\end{tabularx}%
\caption{List of radial cutoff parameters (core radii, in atomic units) for each angular quantum number and default $E_{pw}$ in eV for all potentials employed. 
The GW ultrasoft (GW-US) versions of the PAW potentials were used for all materials except for SrTiO$_3$, which employed the norm-conserving GW (GW-NC) ones.}
\label{tab:CompDetails:POTCAR}
\end{table}
\newline The procedure used to determine the optical conductivity for each material involves three main steps:
\begin{enumerate}[leftmargin=*]
\item The first one is a standard self-consistent DFT calculation using the generalized gradient approximation (GGA) parametrized by Perdew, Burke and Ernzerhof (PBE)~\cite{PBE.PhysRevLett.77.3865}. 
For LaTiO$_3$ and LaVO$_3$ PBE alone is not able to open the gap - therefore a small effective onsite Hubbard U$_{\rm eff}$= 2 eV~\cite{Ergorenc.PRM.2.024601} was added, using the DFT+U formulation of Dudarev~\cite{DFTU.PhysRevB.57.1505}. The spin-orbit coupling (SOC) is included for NaOsO$_3$~\cite{SOC.PhysRevB.93.224425}.
\item The second step consists of a single-shot $G_0W_0$ variant of the $GW$ approximation: in the employed implementation~\cite{G0W0.PhysRevB.74.035101} the one-particle Green's functions are constructed from the previously determined PBE one-electron energies and orbitals, while the dynamically screened Coulomb interaction $W$ is computed within the framework of the RPA~\cite{G0W0.PhysRevB.74.035101, G0W0.PhysRevB.49.8024}.
The convergence of the quasi-particle (QP) energies for this dataset has been previously investigated by Erg{\"o}renc \textit{et al.}~\cite{Ergorenc.PRM.2.024601}. The energy cutoffs, number of bands and numbers of frequency points, summarized in Table~\ref{tab:CompDetails:Param}, are chosen consistently with the setup adopted in Ref.~\cite{Ergorenc.PRM.2.024601} in order to to ensure converged QP gaps within an accuracy of approximately $100$ meV. The choice of the k-point mesh has a paramount impact on the convergence of the optical properties, which will be discussed in details in Section~\ref{sec:CompDetails:k-pointconv}. 
\item The optical conductivities, both in the RPA~\cite{IPA.PhysRevB.73.045112} and in the BSE~\cite{BSE.PhysRevLett.80.4510,BSE.PhysRevLett.81.2312,Sander.PRB.92.045209}, are calculated  from the $G_0W_0$ data. The Tamm-Dancoff approximation (TDA)~\cite{Dancoff.PhysRev.78.382} is used; this approximation has been proven to be reliable for the prediction of optical spectra for standard semiconductors~\cite{Sander.PRB.92.045209}.
\end{enumerate}
The frequency-dependent macroscopic dielectric function is calculated as~\cite{Sander.PRB.92.045209,Liu.PRM.2.075003}
\begin{equation}
\begin{aligned}
\varepsilon(\omega)&=1- \lim _{\mathbf{q} \rightarrow 0} V(\mathbf{q}) \sum_{\Lambda}\left(\frac{1}{\omega-E^{\Lambda}+i \eta}-\frac{1}{\omega+E^{\Lambda}-i \eta}\right) \\
& \times\left\{\sum_{\mathbf{k}} w_{\mathbf{k}} \sum_{v, c}\left\langle\psi_{c \mathbf{k}}\left|e^{i \mathbf{q} \cdot \mathbf{r}}\right| \psi_{v \mathbf{k}}\right\rangle A_{c v \mathbf{k}}^{\Lambda}\right\} \times\{c . c .\},
\end{aligned}
\end{equation}
where $E^\Lambda$ and $A_{c v \mathbf{k}}^{\Lambda}$ are the BSE eigenvalues and eigenvectors, V is the bare Coulomb interaction,  $w_{\mathbf{k}}$ are the k-point weights, $\eta$ a positive infinitesimal and $\psi_{c \mathbf{k}}$ and $\psi_{v \mathbf{k}}$ are respectively the unoccupied and occupied DFT wavefunctions. The oscillator strengths $S_\Lambda$ associated with the optical transitions are defined as~\cite{Liu.PRM.2.075003}
\begin{equation}
  S_\Lambda = Tr \left[\sum_{\mathbf{k}} w_{\mathbf{k}} \sum_{v, c}\left[\langle\psi_{c \mathbf{k}}\left|e^{i \mathbf{q} \cdot \mathbf{r}}\right| \psi_{v \mathbf{k}}\right\rangle A_{c v \mathbf{k}}^{\Lambda} \times\{c . c .\} \right].
  \end{equation}
The exciton binding energies $E_{xb}$ are computed as the difference between the first bright BSE transition and the fundamental $G_0W_0$ gap.
The optical results will be interpreted in Sec.~\ref{sec:CubicPerov:MainOptTrans} in terms of the joint density of states (JDOS) defined as:
\begin{equation*}
 {\rm JDOS}(\omega) = 2 \sum_{v,c,\mathbf{k}} w_{\mathbf{k}} \delta \left(E_c(\mathbf{k})-E_v(\mathbf{k}) -\hbar \omega  \right),   
\end{equation*}
 where $E_c(\mathbf{k})$ and $E_v(\mathbf{k})$ are the $G_0W_0$ eigenvalues; the Dirac $\delta$ is approximated by a normalized Gaussian function with a broadening parameter of $0.10$ eV.
\subsection{\label{sec:CompDetails:k-pointconv}k-point convergence}
\begin{figure*}[!hbt]
	\includegraphics[width=17.6cm]{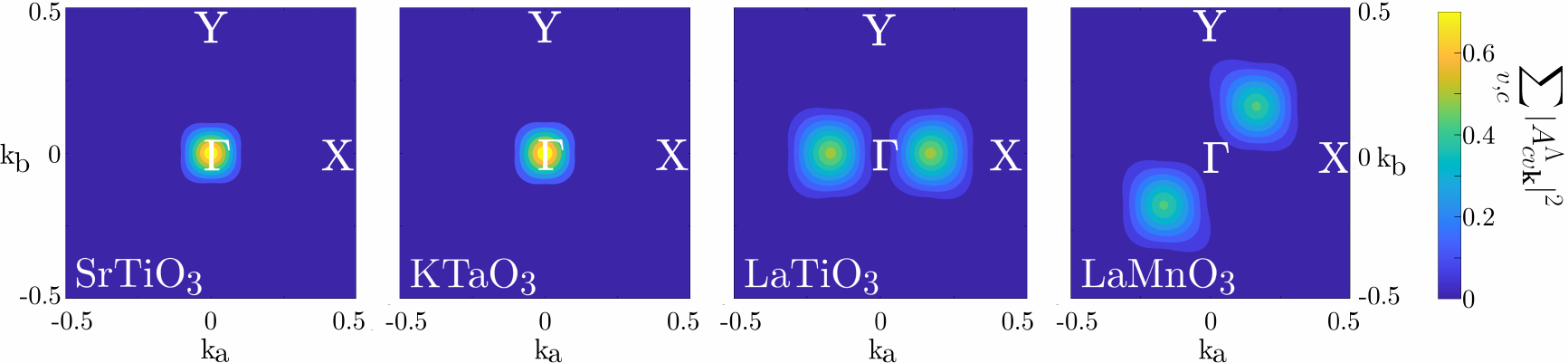} 
\caption{\label{fig:CompDetails:2Dexciton}Contour plots of the squared modulus $\sum_{v,c} \abs{A_{cv \textbf{k}}^\Lambda}^2$ of the first non-dark exciton for SrTiO$_3$, KTaO$_3$, LaTiO$_3$ and LaMnO$_3$ along the $k_a-k_b$ plane ($k_c$=0) in the Brillouin zone. The eigenvectors are calculated with the BSE scheme on a 
$8 \times 8 \times 8$ k-mesh for SrTiO$_3$ and KTaO3$_3$ and on a  $6 \times 6 \times 4$ for LaTiO$_3$ and LaMnO$_3$.}
\end{figure*}

It is well known that optical properties exhibit a strong dependence on k-point sampling and generally very dense k-point meshes are required to obtain well converged optical conductivities~\cite{Albrecht.PRL.83.3971, Kammerlander.PRB.86.125203, Fuchs.PRB.78.085103,Hahn2005, Umari.PRB..103.075201}. It should be noted that in cases where the exciton is spatially highly delocalized it is more efficient to use dense meshes only in a small portion of the Brillouin zone (around $\Gamma$), rather than sampling the full k-space, as recently discussed by P. Umari in Ref.~\cite{Umari.PRB..103.075201}.
In this work, to obtain reliable results we carried out a careful k-point convergence procedure; the convergence criteria employed is based on the accurate quantification of the first BSE eigenvalue $E^\Lambda$ related to a non-dark exciton within an accuracy of 5 meV.
\newline Considering the huge computational cost that a $GW$+BSE calculation on a dense k-point grids may involve, results on dense converged k-meshes as well as the convergence tests themselves {cannot} be efficiently done at the standard $GW$+BSE level.
To mitigate these limitations, two different strategies were adopted: (i) mBSE and (ii) \textit{k}-averaging.
The mBSE scheme is used to perform the convergence tests  and to determine the $E_{xb}$ values, while the \textit{k}-averaging technique is employed to calculate the optical conductivity spectra $\sigma(\omega)$.

\subsubsection{Model-BSE}
The mBSE approach~\cite{Fuchs.PRB.78.085103,  Bokdam.SR.2016.28618} introduces two approximations to the standard BSE scheme:
\begin{enumerate}
    \item Model dielectric screening approximation: the RPA dielectric function calculated in the $G_0W_0$ step is approximated by an analytic model~\cite{Bokdam.SR.2016.28618}
\begin{equation}
    \epsilon^{-1}_{\textbf{G},\textbf{G}}(\textbf{k})= 1 - (1-\epsilon_\infty^{-1})  exp \left[- \frac{\abs{\textbf{k}+\textbf{G}}^2}{4 \lambda^2}\right],
    \label{eq:mbse}
\end{equation}
 where $\epsilon_\infty$ is the static ion-clamped dielectric function ad $\lambda$ the range separation parameter, which is determined by fitting $\epsilon^{-1}_{\textbf{G},\textbf{G}}(\textbf{k})$ to the RPA calculated one. The off-diagonal elements of the inverse dielectric function are neglected rendering the screened Coulomb kernel diagonal ($\textbf{G}=\textbf{G}'$). This analytical model has proven to be a good approximation to the full dielectric function~\cite{Tal.PRResearch.2.032019}. 
 \item The QP energies are approximated through the application of a scissor operator to the DFT one-electron energies (such that the resulting band gap matches the $G_0W_0$ one).
 \end{enumerate}
 
 This approach reduces the overall computational cost and  was successfully applied to halide perovskites~\cite{Bokdam.SR.2016.28618,D0CP00496K}, iridates~\cite{Liu.PRM.2.075003}, 3d TMO~\cite{Liu.JoP.2019}, and it has been shown to correctly reproduce the full BSE spectrum up to 6 eV for SrTiO$_3$~\cite{Begum.PRM.3.065004}. All scissor operators used, along with $\lambda$ and $\epsilon_\infty^{-1}$, are detailed in the Supplementary Materials~\cite{SupplementaryMaterial}.
\begin{table}[]
\begin{tabularx}{\columnwidth}{@{}>{\hsize=1.5cm}X >{\hsize=1.6cm}l >{\hsize=2.6cm}c@{\hspace{0.4cm}}c@{}}
\toprule\toprule
 &\begin{tabular}[c]{@{}l@{}}Converged\\ k-mesh\end{tabular} &
  \begin{tabular}[c]{@{}l@{}}Brillouin zone\\  Volume ($\si{\angstrom}^{-3}$)\end{tabular} &
  \begin{tabular}[c]{@{}l@{}}k-point density \\($kpts/\si{\angstrom}^{-3}$)\end{tabular} \\ \midrule
SrTiO$_3$     & $20 \times 20 \times 20$ & 4.17 & 1920 \\
SrZrO$_3$     & $20 \times 20 \times 20$ & 3.57 & 2240 \\
SrHfO$_3$     & $20 \times 20 \times 20$ & 3.56 & 2240 \\
KTaO$_3$      & $20 \times 20 \times 20$ & 3.91 & 2040 \\
LaScO$_3$     & $10 \times 10 \times 6$  & 0.93 & 640 \\
LaTiO$_3$     & $10 \times 10 \times 6$  & 1.00 & 600 \\
LaVO$_3$      & $10 \times 10 \times 6$  & 1.03 & 580 \\
LaCrO$_3$     & $10 \times 10 \times 6$  & 1.06 & 570 \\
LaMnO$_3$     & $10 \times 10 \times 6$  & 1.02 & 590 \\
LaFeO$_3$     & $10 \times 10 \times 6$  & 1.02 & 590 \\
SrMnO$_3$     & $8\times 8\times 4$      & 1.11 & 230 \\
SrTcO$_3$     & $9\times 9\times 6$      & 1.02 & 480 \\
Ca$_2$RuO$_4$ & $8\times 8\times 4$      & 0.70 & 370 \\
NaOsO$_3$     & $9\times 9\times 6$      & 1.14 & 430 \\ \bottomrule
\end{tabularx}
\caption{Converged k-point grids, volumes of the Brillouin zone (BZ) and k-points densities for the listed materials. 
The k-point density is calculated as the total number of k-points divided by the Brillouin zone volume.
A k-point mesh is considered converged when the BSE eigenvalue related to the first non-dark eigenvector is determined within an accuracy of 5 meV.}
\label{tab:CompDetails:converged_k-point_meshes}
\end{table}
The k-point grids that ensure the required $E_{xb}$ accuracy are presented in Table~\ref{tab:CompDetails:converged_k-point_meshes} and range from $8 \times 8 \times 4$ to $20 \times 20 \times 20$, depending on the system. We note that the cubic (C-P$_{m \bar{3} m}$) perovskites require considerably denser k-meshes than the magnetic compounds.
Nevertheless, even a $20\times20\times20$ mesh does not yield a fully converged $\sigma(\omega)$ over the entire energy range (see for example SrTiO$_3$ and KTaO$_3$ in Fig.~\ref{fig:CompDetails:KptsConvergencExample}).
This is due to the large spatial delocalization of the excitonic wavefunction in these materials, as discussed below.
In contrast, for all remaining perovskites a fully converged $\sigma(\omega)$ has been obtained, and even sparser $k$-point meshes are able to reproduce the spectra (see Fig.~\ref{fig:CompDetails:KptsConvergencExample}).
This behaviour can be traced back to the different degree of localization of the first non-dark BSE eigenvectors $A_{c v \mathbf{k}}^{\Lambda}$, illustrated as contour-plots of the squared modulus $\sum_{v,c} \abs{A_{cv \textbf{k}}^\Lambda}^2$ in \textbf{k}-space for selected examples in Fig.~\ref{fig:CompDetails:2Dexciton}.
 The cubic compounds exhibit an excitonic wavefunction strongly localized around the $\Gamma$ point in the BZ.
 This in turn imposes the necessity of very dense k-point meshes to correctly describe $A_{c v \vec{k}}^\Lambda$~\cite{sottilePhDThesis, Laskowski.PRB.72.035204} and to avoid spurious artificial confinement effects~\cite{Molina.PRB.88.045412}, which can be achieved by selectively increasing the k-point density around $\Gamma$ (this could be achieved within a regular-grid procedure by setting to zero all contributions to the excitonic amplitude beyond a cut-off volume around $\Gamma$~\cite{Umari.PRB..103.075201}).
 The remaining perovskites show excitonic wavefunctions that span a larger portion of the BZ and thus require a less dense BZ sampling.

\subsubsection{\textit{k}-averaging}

A \textit{k}-averaging procedure was adopted to calculate the full BSE spectra~\cite{Sander.PRB.92.045209, Liu.PRM.2.075003}. This averaging procedure includes two steps:
in the first step all $L$ irreducible k-points $\widetilde{\textbf{k}}_{1,..,L}$ from a $\Gamma$-centered $n \times n \times n$ grid are generated; in the second step $L$ independent $GW$+BSE calculations are executed. Each calculation is based on a $m \times m \times m$ grid, shifted by the corresponding $\widetilde{\textbf{k}}_{1,..,L}$.
 The final dielectric function is therefore attained by averaging over the previous results:
 \begin{equation}
     \epsilon(\omega)= \frac{1}{W} \sum_{p=1}^L w_{\widetilde{\textbf{k}}_p} \epsilon_{\widetilde{\textbf{k}}_p}(\omega), \qquad W=\sum_{p=1}^L w_{\widetilde{\textbf{k}}_p}, 
 \end{equation}
where $\epsilon_{\widetilde{\textbf{k}}}$ is the dieletric function calculated  on the mesh shifted by $\widetilde{\textbf{k}}$. The final result, which includes all k-points of a regular $(n\cdot m)\times(n\cdot m)\times(n\cdot m)$ calculation, is denoted by $m\times m \times m | n \times n \times n$.
This \textit{k}-averaging approach implicitly involves an approximation \cite{Sander.PRB.92.045209}: the long-range part of the Coulomb kernel is truncated at $\sim m$ times the unit cell size and consequently may cause spurious artefacts for extended real-space  exciton wavefunctions.
The cubic (C-P$_{m \bar{3} m}$) compounds, in particular, possess a (first non-dark) exciton wavefunction strongly localized around $\Gamma$, which corresponds to a delocalized real-space wavefunction and requires therefore careful testing. The magnetic TMO perovskites are instead less affected by the risk of spurious artefacts due to a more delocalized exciton wavefunction in reciprocal space. 

The employed $n$ and $m$ values were checked by comparing the averaged $\sigma(\omega)$ with a standard BSE calculation without \textit{k}-averaging and investigating the presence of substantial peak enhancements and suppressions.
The choice ($m=4$,$n=5$) does not introduce artificial artifacts in SrHfO$_3$, SrZrO$_3$ and KTaO$_3$; however to avoid a spurious peak suppression inside the SrTiO$_3$ optical spectrum, a larger $m=7$ value is needed (see Supplement Materials~\cite{SupplementaryMaterial} for more details).
\begin{figure*}[!htb]
\centering
\includegraphics[width=15.0cm]{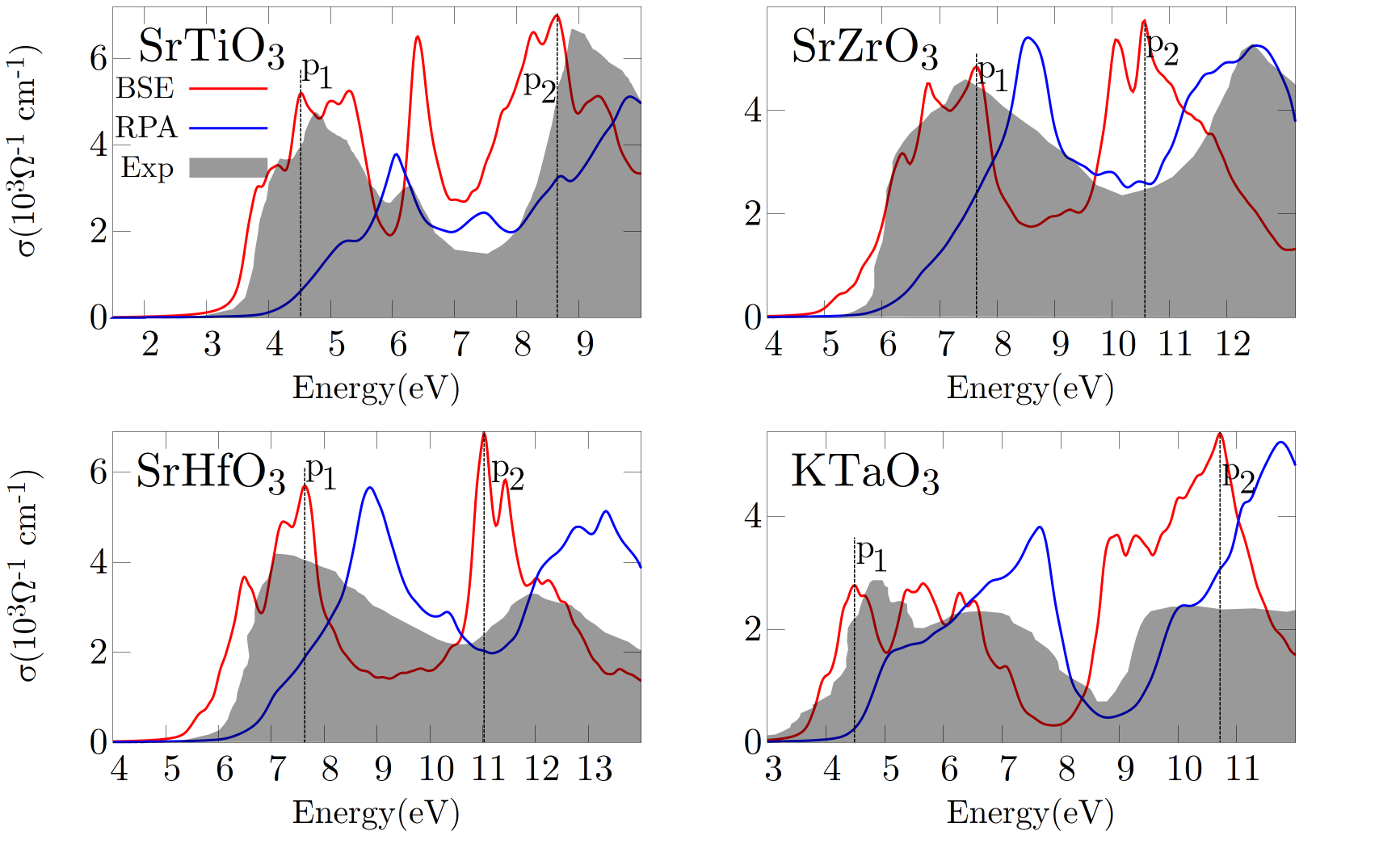}
\caption{\label{fig:CubicNonMagnetic:OptCond}Optical conductivity $\sigma(\omega)$ calculated within the RPA ({\color{blue}blue line}), BSE approach ({\color{red}red line}) and experimental data.
All curves are obtained through an $5\times5\times5 | 4\times4\times4$ \textit{k}-averaging, except for SrTiO$_3$ which employed a $7\times7\times7 | 4\times4\times4$. The main two peaks of the BSE spectra of each structure are labeled as $p_1$ and $p_2$.}
\end{figure*}

\section{Results and Discussion}

The discussion of the results is divided into three sections, each focusing on a specific class of perovskites: (i) cubic non-magnetic perovskites (SrTiO$_3$, SrZrO$_3$, SrHfO$_3$ and KTaO$_3$), (ii) Lanthanum series (LaScO$_3$, LaTiO$_3$, LaVO$_3$, LaCrO$_3$, LaMnO$_3$ and LaFeO$_3$), and (iii) Ca$_2$RuO$_4$, NaOsO$_3$, SrMnO$_3$, and SrTcO$_3$.

\subsection{\label{sec:CubicPerov} Cubic Perovskites}

\subsubsection{\label{sec:CubicPerov:Comparison}Comparison between BSE and RPA spectra}

The optical conductivity $\sigma(\omega)$ obtained through the \textit{k}-averaging procedure for all cubic compounds is shown in Fig.~\ref{fig:CubicNonMagnetic:OptCond}. All spectra exhibit a similar lineshape characterized by two main structures (designated by their most intense peaks $p_1$ and $p_2$) caused by the crystal field splitting of the TM-$d$ states into $t_{2g}$ and $e_g$ subsets.
We note that the very sharp peak observed in SrTiO$_3$ at $\approx$ 6.4 eV has no analogues in the other cubic materials. Its origin has been examined by Ref.~\cite{Sponza.PRB.87.235102,Begum.PRM.3.065004} and has been related to transitions to (low dispersing) localized Ti-$e_g$ states along the $\Gamma-X$ direction in the BZ.
Sponza \textit{et al.}~\cite{Sponza.PRB.87.235102} discussed the neglect of coupling terms (i.e electron-phonon interaction or the dynamical screening) of the standard BSE approach as a possible reason behind the exceedingly strong intensity of the peak, which does not appear in the experimental data.

\begin{figure}[!hb]
        \includegraphics[width=8.6cm]{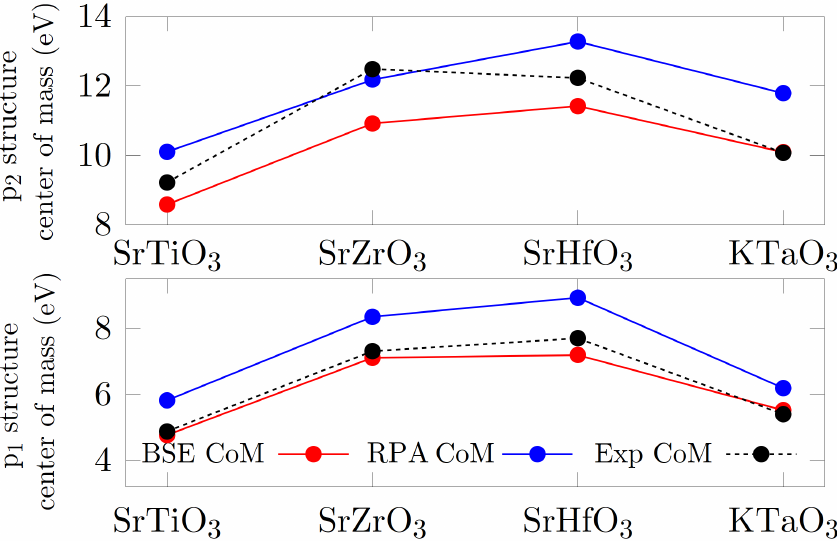}
       \caption{\label{fig:CubicNonMagnetic:CenterOfMasses} Centers of mass (CoM) of the two main structures in the cubic compounds spectra.
       }
\end{figure}

The BSE improves considerably upon RPA the quantitative agreement with the experimental data, in particular for what concern the intensities and energy positions of the first structures.
The differences between the experimental centers of mass (CoM) of the $p_1$ structures and the BSE CoM are strongly reduced, with a mean absolute error of $0.24$ eV compared to $1.00$ eV for the RPA curves (see Fig.~\ref{fig:CubicNonMagnetic:CenterOfMasses}).
Small residual discrepancies between BSE and measured curves are visible at the onset, especially for SrHfO$_3$ and SrZrO$_3$.  Significant contributions to these discrepancies originate from differences between the experimental and $G_0W_0$ predicted gaps~\cite{Ergorenc.PRM.2.024601}, equal to $0.30$ eV (SrTiO$_3$, SrZrO$_3$) and $0.40$ eV (SrHfO$_3$).
The BSE-induced redshift of the $p_1$ structures (evaluated as the difference between the RPA and BSE spectra at the onset at  $\sigma(\omega)\sim 1\times10^3\Omega^{-1}$cm$^{-1}$) varies from 0.80 eV (KTaO$_3$) to 1.20 eV (SrTiO$_3$). Significant spectral weight transfers are thus visible, signaling strong excitonic contributions for the considered cubic systems.

The onsets for the $p_2$ structure are instead systematically underestimated by about 1-2 eV; for SrTiO$_3$ this deviation was attributed to  excessively strong excitonic effects~\cite{Sponza.PRB.87.235102}.
The mean absolute error between the BSE CoM of the $p_2$ structures and the experimental ones (see Fig.~\ref{fig:CubicNonMagnetic:CenterOfMasses}) is equal to $-0.75$ eV. 

The exciton binding energies $E_{xb}$ for different k-meshes are listed in Table~\ref{tab:CubicPerov:Exb}: the converged $E_{xb}$ range between $\approx$ 150 and 250 meV. The use of mBSE for the $E_{xb}$ estimation is justified by a direct comparison with the BSE prediction on a reduced $11 \times 11 \times 11$ mesh. In general the BSE reference values are very well reproduced by the mBSE, with an error varying from $4\%$ (SrZrO$_3$) to $8\%$ (SrHfO$_3$). Our BSE calculated $E_{xb}$ for SrTiO$_3$ (205 meV) is consistent with previous BSE predictions of Begum \textit{et al.}~\cite{Begum.PRM.3.065004} (246 meV) and Sponza \textit{et al.}~\cite{Sponza.PRB.87.235102} (220 meV).
The difference with Begum's result is due to the use of the SCAN functional in Ref.~\cite{Begum.PRM.3.065004}. Increasing the k-point mesh up to  $20 \times 20 \times 20$ within the mBSE leads to a lowering of  $E_{xb}$ to 165 meV.
We note that the choice of the  k-point mesh has a paramount effect on the final values; the change of $E_{xb}$  between the $11\times11\times11$ and the well converged $20\times20\times20$ mesh is between $0.056$ eV (SrZrO$_3$) and $0.088$ eV (KTaO$_3$), see Table~\ref{tab:CubicPerov:Exb}.

\begin{table}[h]
    \begin{tabularx}{\columnwidth}{@{}lcXXXX@{}}
    \toprule\toprule
                                                                          & k-mesh                   & SrTiO$_3$ & SrZrO$_3$ & SrHfO$_3$ & KTaO$_3$ \\ \midrule
    BSE                                                                   & $11 \times 11 \times 11$ & 0.205     & 0.321     & 0.319     & 0.230    \\
    mBSE                                                                  & $11 \times 11 \times 11$ & 0.195     & 0.308     & 0.293     & 0.215    \\ 
     \% error                                                             &                          & 5         & 4         & 8         & 7     \\  \arrayrulecolor{black!45}\cmidrule{2-6}
    mBSE                                                                  & $20 \times 20 \times 20$ & 0.149     & 0.275     & 0.258     & 0.160    \\  \arrayrulecolor{black!45}\midrule
    $\epsilon_\infty^{-1}$                                                &                          & 0.165     & 0.231     & 0.242     & 0.195    \\
    $\lambda$                                                             &                          & 1.463     & 1.457     & 1.448     & 1.420    \\ \arrayrulecolor{black}\bottomrule\bottomrule
    \end{tabularx}
    \caption{\label{tab:CubicPerov:Exb} Exciton binding energies $E_{xb}$ in eV for the cubic materials, calculated through the BSE and mBSE approaches. The third row summarizes the percentage errors between the BSE reference values and the mBSE ones.
    The employed k-point meshes  are specified in the second column ($20 \times 20 \times 20$ represents the converged mesh as described in Section~\ref{sec:CompDetails:k-pointconv}). The calculated inverse static dielectric constants and screening length parameters $\lambda$ ($\si{\angstrom}^{-1}$) used for the mBSE (Eq.~\eqref{eq:mbse}) are given.}
\end{table}
\begin{figure*}[!t]
   	\includegraphics[width=\linewidth]{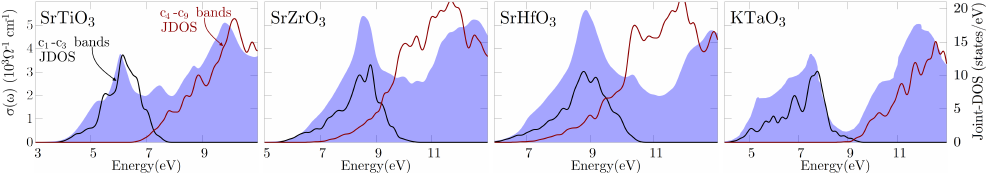}  
	\caption{\label{fig:CubicNonMagnetic:JDOS}Optical conductivity $\sigma(\omega)$ in the RPA ({\color{blue}blue shadow background}), along with the Joint Density of States associated with transitions to the first three conduction bands $c_1-c_3$ (black lines) and to conduction bands $c_4-c_9$ ({\color{red} red lines}). All curves are obtained with an $11\times11\times$ k-point mesh. }
\end{figure*}

\subsubsection{\label{sec:CubicPerov:MainOptTrans}Origin and character of main optical transitions}

\begin{figure*}[p]
\centering
\includegraphics[width=0.75\textwidth]{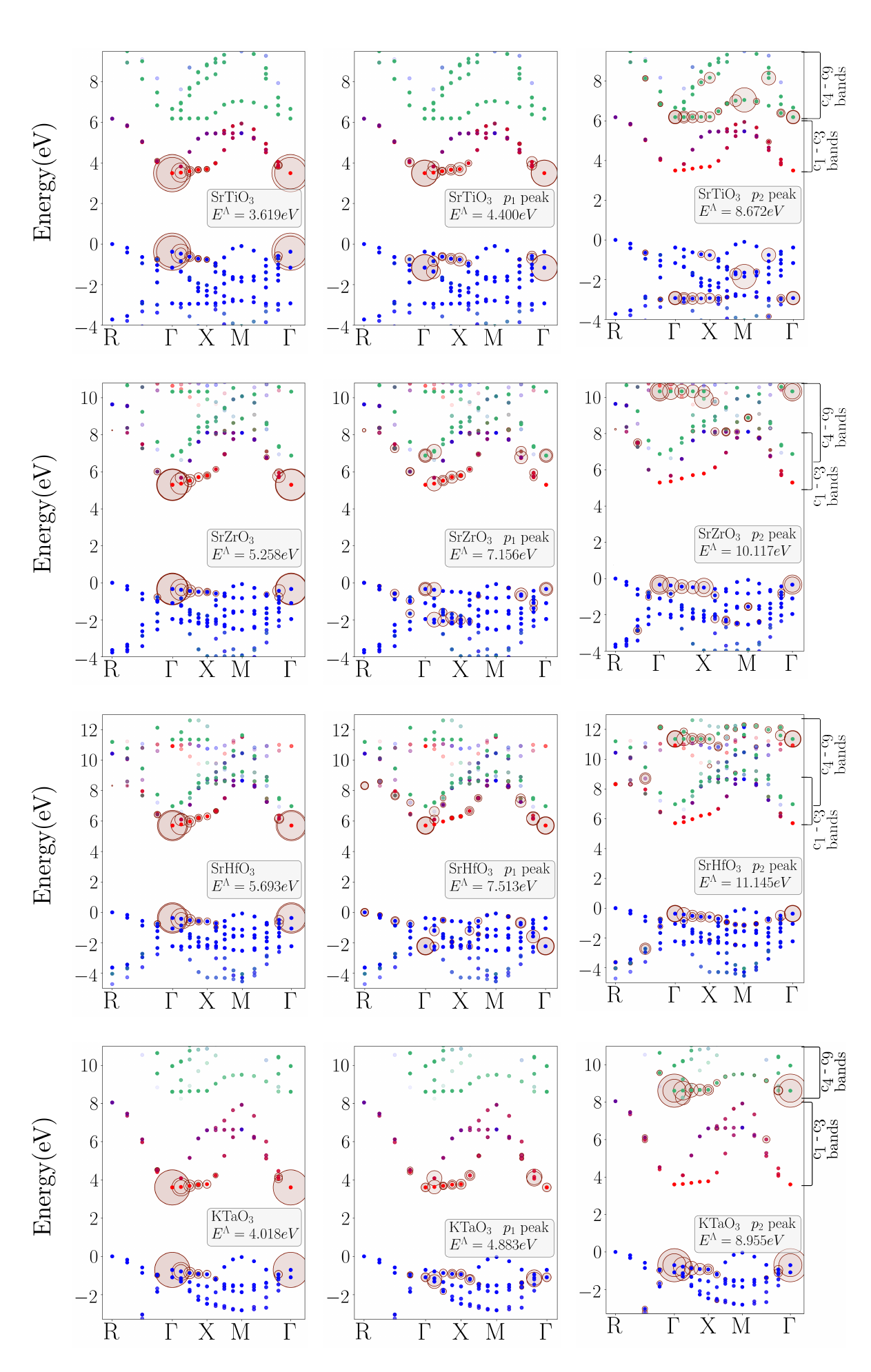}
\caption{\label{fig:CubicNonMagnetic:Fatband}
Fat band pictures for the cubic systems, obtained through $G_0W_0$+BSE on an $8\times8\times8$ k-point grid. 
The circle radius corresponds to the contribution $\abs{A_{ \mathbf{k} v c}^\Lambda}^2$ to the e-h pair wavefunction at that k-point. 
The left panels refer to the \textit{first} non-dark eigenvector of each material; the middle panels refer to the eigenvectors associated with the highest oscillator strengths close to the $p_1$ peaks, and the right panels to the eigenvectors with the highest oscillator strengths close to the $p_2$ peaks.
The colors of the points in the band-structures are associated with different orbital characters: {\color{blue}blue for O$-p$}, {\color{red}red for TM$-t_{2g}$} and {\color{green}green for TM$-e_g$}.}
\end{figure*}

The octahedral crystal field that causes the splitting of the $d$ states is the dominant factor for the formation of the observed two-structure spectra. In order to analyze the character of the transitions we display the JDOS in Fig.~\ref{fig:CubicNonMagnetic:JDOS} (which provides a measure of the number of allowed optical transitions between initial and final states).
Here and in the following the discussion of the optical transitions is given in terms of the band labelling shown in Fig.~\ref{fig:CubicNonMagnetic:Fatband}, where the conduction bands are denoted as $c_n$ with $n=1$ for the first conduction band and so on.

For KTaO$_3$ and SrTiO$_3$ the first structures (around peak $p_1$) are almost completely determined by transitions from the occupied oxygen manifold to the first three conduction bands (denoted as $c_1-c_3$), i.e. $O\text{-}2p \rightarrow c_1-c_3$ (the higher $c_4-c_9$ conduction bands almost do not contribute to the first structures' JDOS).  
For these two compounds the $c_1-c_3$ manifolds have a main TM$-t_{2g}$ character, with a limited O$-p$ hybridization away from the $\Gamma$ point (between 5$\%$ and 15$\%$ for SrTiO$_3$ and between 2$\%$ and 23$\%$ for KTaO$_3$, as determined by $G_0W_0$ calculations on an $8\times8\times8$ k-meshes).

The analysis of SrHfO$_3$ and SrZrO$_3$ optical transitions is complicated by the presence of band entanglement between 
the $c_1-c_3$ manifold and the upper $c_4-c_9$ manifold, not present in KTaO$_3$ and SrTiO$_3$ (where these two sets of bands are separated in energy, see Fig.~\ref{fig:CubicNonMagnetic:Fatband}).
Similarly to SrTiO$_3$ and KTaO$_3$, the $c_1-c_3$ bands possess a main TM$-t_{2g}$ character, which however is not uniform in the Brillouin zone and exhibits a significant hybridization with O$-p$ states (between $3\%$ and $27\%$ for SrZrO$_3$ and between $2\%$ and $31\%$ for SrZrO$_3$, as determined by $G_0W_0$ calculations on $8\times8\times8$ k-meshes).
Due to this band-entanglement the JDOS shows some contributions to the $p_1$ structures from transitions to bands $c_4-c_9$ (see Fig.~\ref{fig:CubicNonMagnetic:JDOS}).
To ascertain the actual importance of these transitions and the relative significance of the $c_1-c_3$ and $c_4-c_9$ more quantitatively  we list in Table~\ref{tab:CubicPerov:S1S2} the BSE eigenvectors relative to the $p_1$ peak for each material in terms of the 
amplitude distribution (i.e. the total square amplitude) associated with transitions to bands $c_1-c_3$
($D_{c_1-c_3}^\Lambda\hspace{-0.1cm}=\sum_k\sum_{v\in O-2p} \sum_{c\in c_1-c_3} \abs{A_{ \mathbf{k} v c}^\Lambda}^2$) and bands $c_4-c_9$ ($D_{c_1-4_9}^\Lambda\hspace{-0.1cm}=\sum_k\sum_{v\in O-2p} \sum_{c\in c_4-c_9} \abs{A_{ \mathbf{k} v c}^\Lambda}^2$).
{The eigenvalues listed in Table~\ref{tab:CubicPerov:S1S2} are chosen as the transitions with the highest oscillator strength close to the $p_1$ peaks.}
Recalling that  $A_{ \mathbf{k} v c}^\Lambda$ is normalized ($\sum_{\mathbf{k},v,c} \abs{A_{ \mathbf{k} v c}^\Lambda}^2=1)$,
the data in Table~\ref{tab:CubicPerov:S1S2} indicate that transitions to $c_1-c_3$ provide  $\sim 90\%$ of the total spectral weight in SrZrO$_3$ and SrHfO$_3$, about $ 9\%$ lower than the corresponding amplitudes in SrTiO$_3$ and KTaO$_3$.

\begin{table}[h]
\begin{tabularx}{\columnwidth}{@{}>{\hsize=2.4cm}X XXXXXX@{}}
\toprule\toprule
	 & SrTiO$_3$       & SrZrO$_3$ & SrHfO$_3$ & KTaO$_3$    \\ \midrule
	$D_{c_1-c_3}^\Lambda$     & 0.99      & 0.89      & 0.88      & 0.99     \\
	$D_{c_4-c_9}^\Lambda$     & 0.01      & 0.11      & 0.12      & 0.01     \\
	$E^\Lambda$ (eV)          & 4.400     & 7.178     & 7.513     & 4.883    \\ \bottomrule\bottomrule
\end{tabularx}
\caption{\label{tab:CubicPerov:S1S2} 
Comparison between BSE amplitude distributions $D^\Lambda$ related respectively to the final states $c_1-c_3$ ($D_{c_1-c_3}^\Lambda\hspace{-0.1cm}=\sum_k\sum_{v\in O-2p} \sum_{c\in c_1-c_3} \abs{A_{ \mathbf{k} v c}^\Lambda}^2$) and $c_4-c_9$ ($D_{c_1-4_9}^\Lambda\hspace{-0.1cm}=\sum_k\sum_{v\in O-2p} \sum_{c\in c_4-c_9} \abs{A_{ \mathbf{k} v c}^\Lambda}^2$). \\$E^\Lambda$ represent the corresponding BSE eigenvalues: the analyzed transitions are associated with the $p_1$ peaks. The data are obtained using an $11\times11\times11$ k-point mesh.
}
\end{table}

\begin{figure*}[t!]
\centering
\includegraphics[width=\textwidth]{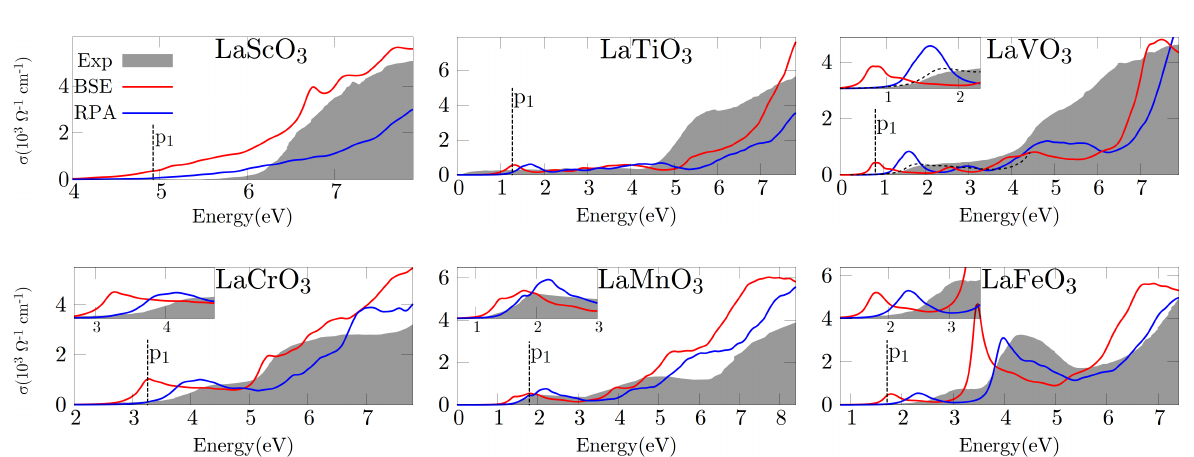}
\caption{\label{fig:LaSeries:OptCond} $\sigma(\omega)$ within the RPA ({\color{blue}blue line}), BSE approach ({\color{red}red line}) and experimental data (from Arima \textit{et al.}~\cite{Arima.PRB.48.17006};  the dashed line for LaVO$_3$ represents the optical conductivity measurement from Miyasaka \textit{et al.}~\cite{Miyasaka.JPSJ.71.2086} ). All curves are obtained through a $5\times2\times5 | 2\times2\times2$ \textit{k}-averaging, except for LaScO$_3$, for which a $5\times5\times3 | 2\times2\times2$ mesh was used. The $p_1$ labels denote the first (low intensity) BSE structures.}
\end{figure*}

Additional insights on the character of the dominant optical transitions  can be extracted by the fat band analysis of the BSE eigenvectors. 
{This is displayed in Fig.~\ref{fig:CubicNonMagnetic:Fatband} for the first non-dark excitons as well as for  $A^\Lambda_{v c k}$ with the highest oscillator strength close to the $p_1$ peaks and $p_2$ peaks.}
As a general feature, common to all cubic materials, the contributions to $\abs{A^\Lambda_{v c k}}$ are predominantly localized at $\Gamma$ (in particular for the first excitations, from the top of the valence band to the bottom of the conduction band) and, less intensively, along the $\Gamma-X$ direction. 
The fat band plots support the association of  the $p_1$ peak with the transitions from the occupied $O-p$ bands (blue) to $t_{2g}$ (red, $c_1-c_3$).
{These features hold not only for the  $A^\Lambda_{v c k}$ related to the $p_1$ transitions, but more generally also for all bright eigenvectors in a $\sim 0.3$ eV range around these $p_1$ transitions.}

\subsection{\label{sec:LaPerov} L\lowercase{a} series}

\subsubsection{\label{sec:LaPerov:Comparison}Comparison between BSE and RPA spectra}

The optical conductivity spectra for the La-based perovskites are collected in Fig.~\ref{fig:LaSeries:OptCond}. All members of the La series exhibit a qualitatively similar $\sigma(\omega)$ (with the exception of LaScO$_3$), characterized by two main different features: a low intensity structure at low energies and a second peak at higher energies, broader and more intense.
The low intensity peak is mainly associated with a Mott-Hubbard type (MH) $d$-$d$ fundamental gap, while the second is typically associated with a charge-transfer (CT) type gap~\cite{Arima.PRB.48.17006, He.PRB.86.235117, Ergorenc.PRM.2.024601}. LaCrO$_3$ in particular can be better described by a mixed MH/CT state, where the first peak is essentially merged with the CT transitions~\cite{He.PRB.86.235117}. The band insulator LaScO$_3$ does not obviously show any Mott-like $d-d$ transition.\\
The transition energies related to the first excitation are well reproduced already at the RPA level.
The systematic  redshifts produced by the excitonic effects (evaluated at $\sigma(\omega)\sim 0.3\times10^3\Omega^{-1}$cm$^{-1}$) are reduced compared to the cubic compounds and vary from $0.3$ eV (for LaTiO$_3$ and LaMnO$_3$) to $0.5$ eV (LaCrO$_3$ and LaFeO$_3$) and $0.7$ eV (LaVO$_3$). Therefore, when compared to the experimental curves, the BSE approach produces an underestimation of the first transition energies for almost all compounds. 
The only exception is LaTiO$_3$, where the optical gap is overestimated as a consequence of the corresponding overestimation of the experimental gap ($0.1$ eV)~\cite{Arima.PRB.48.17006}  obtained at $G_0W_0$ level ($\approx$ 0.5 eV, as  described by the non-extrapolated case of~\cite{Ergorenc.PRM.2.024601}). 
The sources of the above deviations are discussed below for each compound. This involves both theoretical arguments and aspects of the experimental measurements (for instance, the available experimental data were obtained by different techniques at different temperatures, making a consistent comparison with  computational data achieved at 0 K difficult, see Table S1 in the SM).
The second structure, located at $7-8$ eV, dominates the spectra and exhibits stronger excitonic effects, with redshifts (evaluated at $\sigma(\omega)\sim 3\times10^3\Omega^{-1}$cm$^{-1}$) between $0.6$ eV (LaCrO$_3$ and LaFeO$_3$) and $1.0$ eV (LaVO$_3$). An improvement over the RPA is observed only for LaTiO$_3$, LaVO$_3$ and LaCrO$_3$; the quality of the agreement with the experimental curves is overall material dependent.

\begin{figure}[!ht]
	\includegraphics[width=\columnwidth]{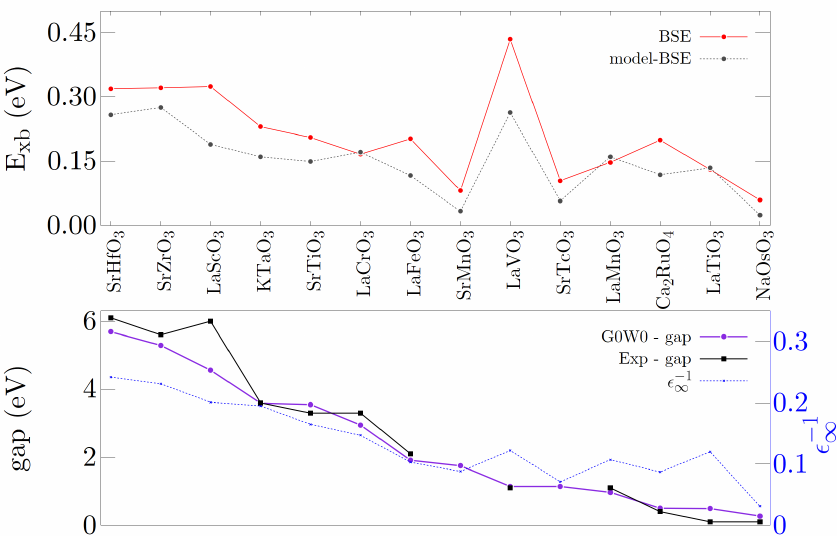} 
	\caption{
	\label{fig:ExbTrend}	Comparison between BSE and mBSE $E_{xb}$ values calculated on the non-converged k-meshes (listed in Tables~\ref{tab:CubicPerov:Exb},~\ref{tab:La:Exb} and ~\ref{tab:RemainingPerv:Exb}).
	 The lower graph displays the $G_0W_0$ and experimental gaps along with the inverse static dielectric constants.
	}
\end{figure}

The ($p$$\rightarrow$$d$) band insulator LaScO$_3$ follows a trend dissimilar to the picture described above: it presents a single, wide and intense peak with a rather strong excitonic redshift of $0.9$ eV (evaluated at $\sigma(\omega)\sim0.3\times10^3\Omega^{-1}$cm$^{-1}$). The disagreement between the measured and calculated spectrum should be traced back to the difference between the $G_0W_0$ and the experimental gap ($\approx$ 1 eV~\cite{Ergorenc.PRM.2.024601}), which has been attributed 
to difficulties in measuring the long tail in the bottom part of the spectrum~\cite{Ergorenc.PRM.2.024601}.
\newline The exciton binding energies (listed in Table~\ref{tab:La:Exb}) fall within the 120 meV - 190 meV range and are overall smaller than their cubic non-magnetic counterparts (with the exception of LaVO$_3$). The predicted higher $E_{xb}$ for LaVO$_3$ is consistent with the larger experimental value of Lovinger \textit{et al.}~\cite{PRB.Lovinger} ($\sim0.6$ eV).
Recent experimental optical conductivity measurements~\cite{Miyasaka.JPSJ.71.2086} also highlight a splitting of the low-energy structure (visible as an additional shoulder at lower temperatures) which has been related to excitonic effects~\cite{PRB.Lovinger,PRB.Kim,PRB.Reul} and is correctly reproduced by the BSE data (but completely absent in the RPA curve).
\newline Compared to the full BSE, mBSE introduces an error ranging between $10\%$ (LaScO$_3$) and $37\%$ (LaVO$_3$), with a mean absolute error of $0.20$ eV. The mBSE scheme therefore performs less satisfactorily for this subset than for the cubic non-magnetic perovskites; {LaVO$_3$ exhibit the larger discrepancy both in absolute and percentage values of the whole set, as shown in Fig.~\ref{fig:ExbTrend}, where we collect the calculated (m)BSE  excitonic binding energies, quasi particles G$_0$W$_0$ gaps and static dielectric constant for all materials included in the considered perovskites dataset}. To gain insight on the cause of the larger error observed for LaVO$_3$, we performed a mBSE calculation on top of the $G_0W_0$ band structure (while keeping the k-mesh, $\mu$ and $\lambda$ fixed at the values of Table~\ref{tab:La:Exb}). In this manner the scissor operator is not required and we can isolate and gauge the effect of the model dielectric function approximation alone. The resulting mBSE@$G_0W_0$ exciton binding energy is only slightly increased with respect to the mBSE value ($0.299$ eV vs. $0.273$ eV), and still much smaller than $E_{xb}^{BSE} = 0.434$~eV, implying that the disagreement mostly arises from the model dielectric function approximation.
{To verify this hypothesis we have conducted an additional BSE@$G_0W_0$ calculation retaining only the diagonal elements of the screened exchange kernel. The resulting binding energy is $E_{xb}^{BSE-diag\,only}=0.322$ eV, much closer to the mBSE@G0W0 value of $0.299$ eV than to $E_{xb}^{BSE}$ with the full screened exchange kernel ($0.434$~eV). This proves that including off-diagonal elements in the inverse dielectric function and in the screened kernel is essential for accurately describing the excitonic properties of LaVO$_3$ and explains the limits of the model dielectric screening approximation for this material.}

\begin{table}[h]
    \centering
    \begin{tabularx}{\columnwidth}{@{}lcXXXXXX@{}}
    \toprule\toprule
                           & \multirow{2}{*}{\begin{tabular}[c]{@{}c@{}}k-point\\ mesh\end{tabular}} & \multicolumn{6}{c}{LaTMO$_3$} \\ \arrayrulecolor{black}\cmidrule(l){3-8} 
                           &                                                                        & Sc    & Ti    & V     & Cr    & Mn    & Fe    \\ \midrule
    BSE                    & $6\times6\times4$                                                      & 0.324 & 0.130 & 0.434 & 0.166 & 0.147 & 0.202 \\
    mBSE                   & $6\times6\times4$                                                      & 0.292 & 0.145 & 0.273 & 0.200 & 0.181 & 0.162 \\
   \% error      &                                                                                  & 10  & 12  & 37  & 20  & 23  & 20   \\ \arrayrulecolor{black!45}\cmidrule{2-8}
    mBSE                   & $10\times10\times6$                                                    & 0.189 & 0.134 & 0.263 & 0.171 & 0.160 & 0.116 \\ \arrayrulecolor{black!45}\midrule
    $\epsilon_\infty^{-1}$ &                                                                        & 0.201 & 0.120 & 0.122 & 0.147 & 0.107 & 0.103 \\
    $\lambda$              &                                                                        & 1.462 & 1.349 & 1.420 & 1.393 & 1.335 & 1.336 \\ \arrayrulecolor{black}\bottomrule\bottomrule
    \end{tabularx}
    \caption{\label{tab:La:Exb}Exciton binding energies $E_{xb}$ in eV for the La series compounds, calculated by mBSE and BSE approaches. Conventions used are the same as in Table~\ref{tab:CubicPerov:Exb}. 
    }
\end{table}

\begin{figure*}[t!]
\centering
\includegraphics[width=\textwidth]{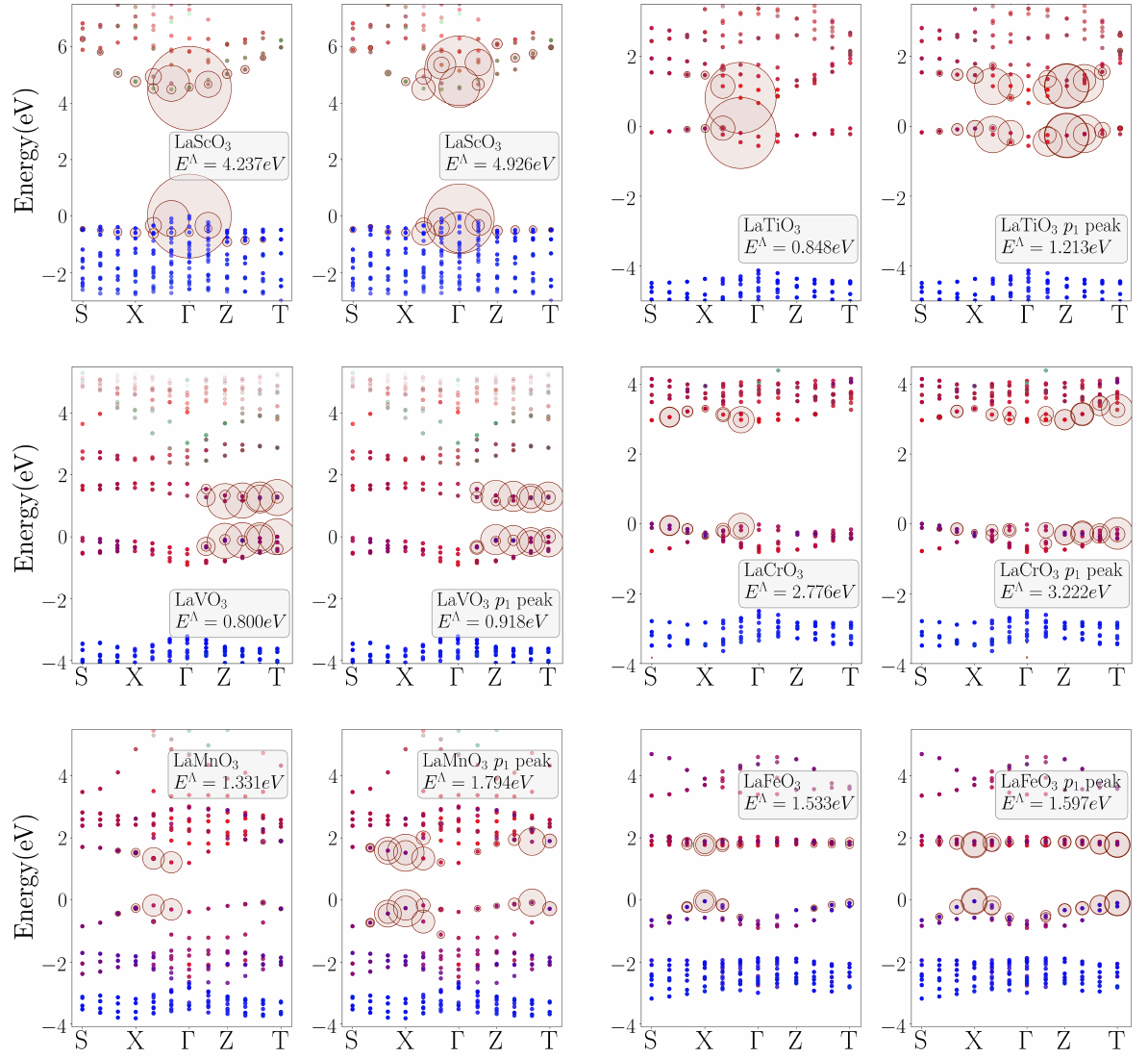}
\caption{\label{fig:LaSeries:Fatbanb}
Fat band plots for members of the La series (LaTiO$_3$, LaVO$_3$, LaCrO$_3$, LaMnO$_3$, LaFeO$_3$ and LaScO$_3$).
For each compound we show two panels: the left one refers to the \textit{first} non-dark eigenvector whereas the  right one is related to the eigenvector with the highest oscillator strength close to the $p_1$ BSE structures (see Fig.~\ref{fig:LaSeries:OptCond}).
Color codings and labellings are  similar to those adopted in Fig.~\ref{fig:CubicNonMagnetic:Fatband}: {\color{blue}blue for O$-p$}, {\color{red}red for TM$-d$} and {\color{green}green for La$-d$}.}
\end{figure*}

\subsubsection{\label{sec:LaPerov:MainOptTrans}Origin and character of main optical transitions}

Also for this series of compounds, we decode the character of the optical transition by analysing the excitonic eigenvectors in a fat band mode. 
Figure~\ref{fig:LaSeries:Fatbanb} presents the fat band pictures related to the first non-dark excitation and to the most intense oscillator strength of the $p_1$ peaks (as indicated in Fig.~\ref{sec:LaPerov:Comparison}). 
The character of the excitonic wavefunction is closely connected to the electronic nature of the insulating state (band insulator, MH, CT and mixed MH/CT)~\cite{He.PRB.86.235117, Arima.PRB.48.17006}.

For the band-insulator LaScO$_3$ (top-left) the direct transitions at $\Gamma$ dominate the excitonic wavefunctions (only minor contributions can be seen along the $\Gamma-X$ direction) which involve $O-p$ to $Sc-d$ excitations.
The first (lowest) set  of optical excitations for MH insulators LaTiO$_3$ (top-right), LaVO$_3$ (middle-left) and LaMnO$_3$ (bottom-left) are determined by $d$-$d$ transitions exclusively involving the two MH sub-bands, whose states have a predominant TM$-d$ character~\cite{He.PRB.86.235117, Nohara.PRB.79.195110, Varignon.Nat.10.1658, Bongjae.PRB.98.075130}. We note that contributions to eigenstates $A^\Lambda_{v c \textbf{k}}$ at $\Gamma$ are almost negligible. This can be explained by recalling that $d \rightarrow d$ transitions are dipole forbidden at k-points with a small point group equal to the full point group of the crystal, like the $\Gamma$ point~\cite{Rodl.PRB.86.235122, Bechstedt.book}.  However, the remaining region of the BZ has a small point group with a lowered symmetry, thus allowing the $d \rightarrow d$ transitions determining the Mott peaks. 
The second main structures (for energies approximately larger than 4 eV) are instead determined by $p-d$ transitions from the valence O$-p$ bands (laying below the occupied MH subband) to the conduction MH subband; at higher energies transitions to La$-d$ states are also involved (not shown).

\begin{figure}[hb]
	\includegraphics[height=7.0cm]{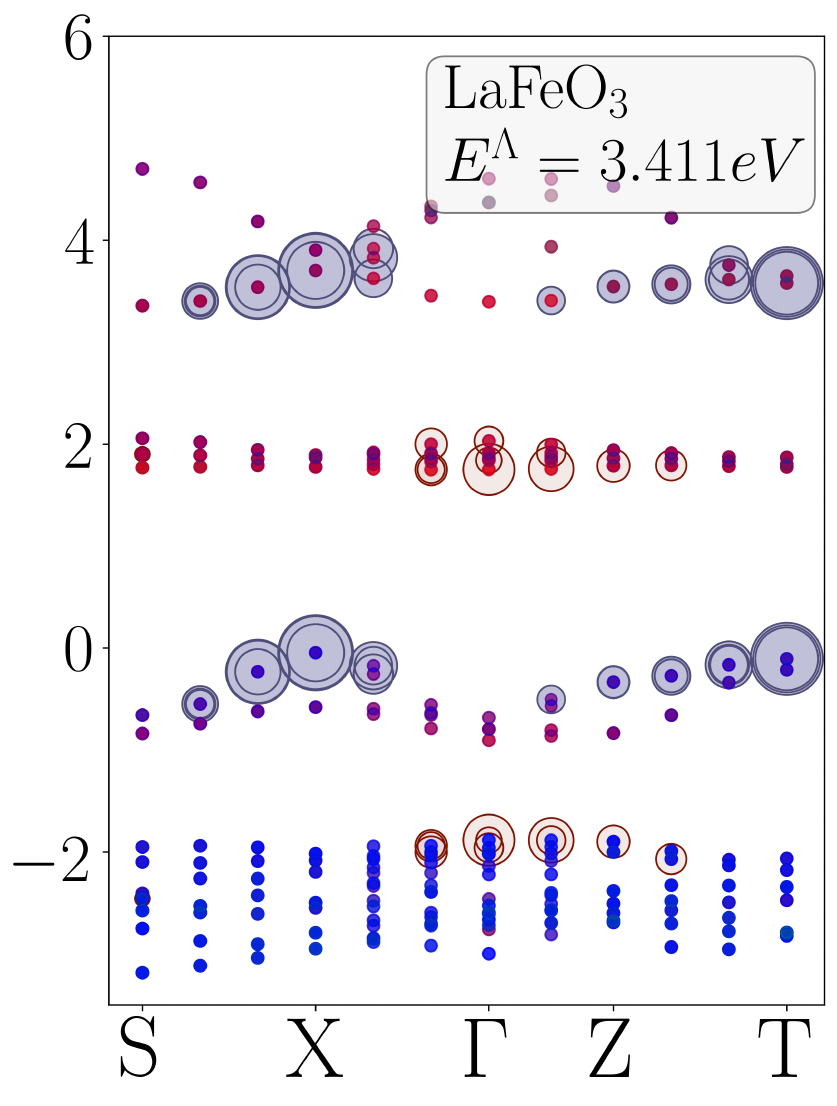} 
	\caption{\label{fig:LFO:FatbandPicture}Fat band picture for $A_{ \mathbf{k} v c}$ associated with the third peak of LaFeO$_3$. The two main transition categories contributing to the eigenvector are distinguished by different colors. }
\end{figure}

For LaCrO$_3$ and LaFeO$_3$ optical experiments reported the coexistence of MH/CT-type excitations at the fundamental gap~\cite{Arima.PRB.48.17006}. This was later confirmed by first principles analyses which indicate a sizable admixture of O-$p$ ($\approx 20 \%/30 \%$ for LaCrO$_3$ and LaFeO$_3$ respectively) and TM-$d$ ($\approx 80 \%/70 \%$)~\cite{Ergorenc.PRM.2.024601,Nohara.PRB.79.195110,He.PRB.86.235117,Yang.PRB.60.15674}.
This mixed CT/MH nature of the optical excitations in LaFeO$_3$ is well captured by the BSE eigenvectors shown in the fat band plots of  Fig.~\ref{fig:LaSeries:Fatbanb}.  
Moreover, the optical spectrum of LaFeO$_3$ exhibits a peculiar third intense peak at $\sim 3-4$eV, 
whose contributions are analyzed in Figure~\ref{fig:LFO:FatbandPicture}. 
Transitions from the mixed O$-p/$Fe$-d$ subband near the Fermi energy to the Fe$-d$ states located at $\sim 4$ eV  provide the majority of the total square amplitude ($\sum_{\mathbf{k}}\sum_{v \in O\text{-}p/Fe\text{-}d}\sum_{c \in Fe\text{-}d\sim 4eV}$ $\abs{A_{ \mathbf{k} v c}^\Lambda}^2 \sim 0.59$) and are particularly intense at the $X$ and $T$ points. A  secondary contribution emerges from valence O$-p$ states at $\sim -2$ eV to the conduction bands at $\sim 2$ eV (with a $\sim 30\%$ total square amplitude).

\begin{figure}[t!]
	\includegraphics[height=7.9cm]{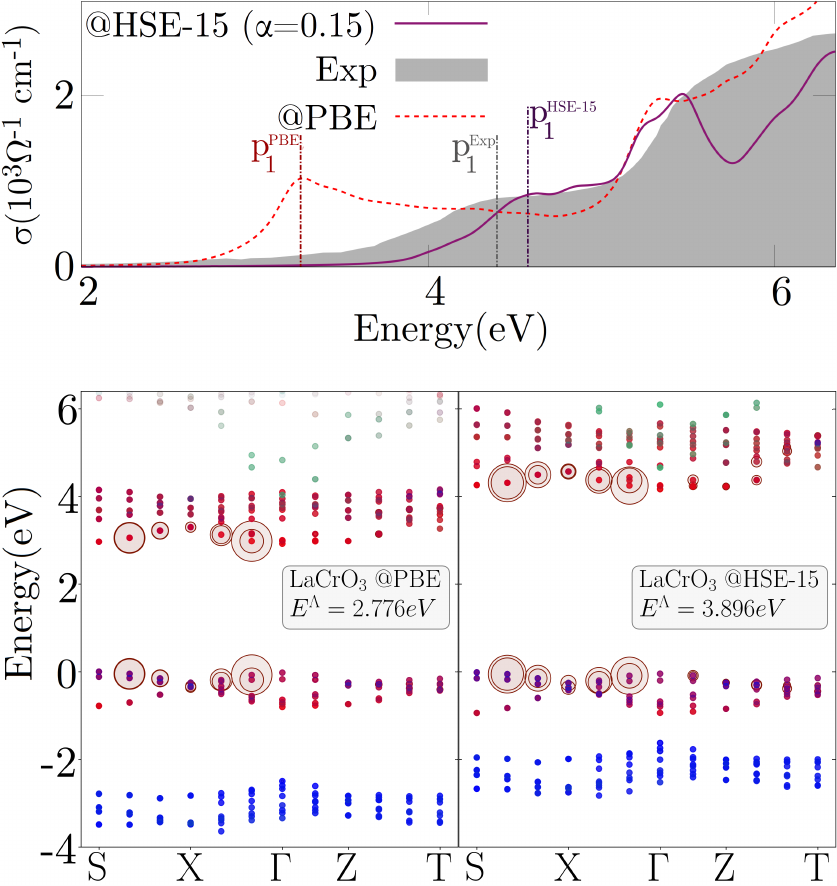} 
	\caption{\label{fig:LCO:SpectraFatbandPicture}
	Comparison between LaCrO$_3$ $G_0W_0$+BSE spectra and fat band pictures calculated from the PBE functional and from the hybrid functional (with an exchange fraction of 0.15). The $p_1$ peak of each spectrum is labeled. The fat band pictures are associated with the first non-dark transitions.}
\end{figure}

For LaCrO$_3$ the coexistence of MH/CT-type transitions at the optical gap is associated with an overlapping of the Mott and CT excitations in the spectrum~\cite{Arima.PRB.48.17006} and has been explained in terms of a significant mixing of Cr-$t_{2g}$ and O-$p$ at the valence band top~\cite{Nohara.PRB.79.195110,He.PRB.86.235117,Yang.PRB.60.15674}.
The optical conductivity in Figure~\ref{fig:LaSeries:OptCond} only partially agrees with this picture:  the energy separation between the $p_1$ peak and the CT structure is significantly overestimated  and the optical gap possesses a dominant $d-d$ character. 
This reduced mixed CT/MH character is due to a low O-$p$ orbital character of the LaCrO$_3$ valence band maxima~\cite{Ergorenc.PRM.2.024601} (with an O-$p$ percentage of $17\% - 20\%$).
\newline Considering the perturbative nature of the $G_0W_0$ scheme, this apparent discrepancy could originate from the PBE starting point. To test this hypothesis we performed an additional $G_0W_0$+BSE calculation starting from hybrid functional orbitals (following the setup of Ref.~\cite{He.PRB.86.235117}, with an exchange fraction $\alpha=0.15$).  The results, shown in Figure~\ref{fig:LCO:SpectraFatbandPicture}, lead to an improved agreement with the experimental data. The oxygen character of the top of the valence band increases from $\sim 20\%$ to $\sim 30\%$, restoring the MH/CT mixed nature of the optical gap. The O-$p$ valence bands below the MH subband are shifted towards higher energies, producing a reduction of the energy separation between the $p_1$ and CT peaks. {However $G_0W_0$ on top of Heyd-Scuseria-Ernzerhof (HSE) hybrid functional overestimates the experimental optical gap by $\sim$ 0.6 eV (with BSE optical gap of  3.89 eV versus the experimental value of 3.30 eV~\cite{Arima.PRB.48.17006}). Conversely, the calculations based on $G_0W_0$@PBE (on the same k-mesh) predicts a smaller optical gap of 2.74 eV. 

\begin{figure*}[!htb]
\centering
\includegraphics[width=15.0cm]{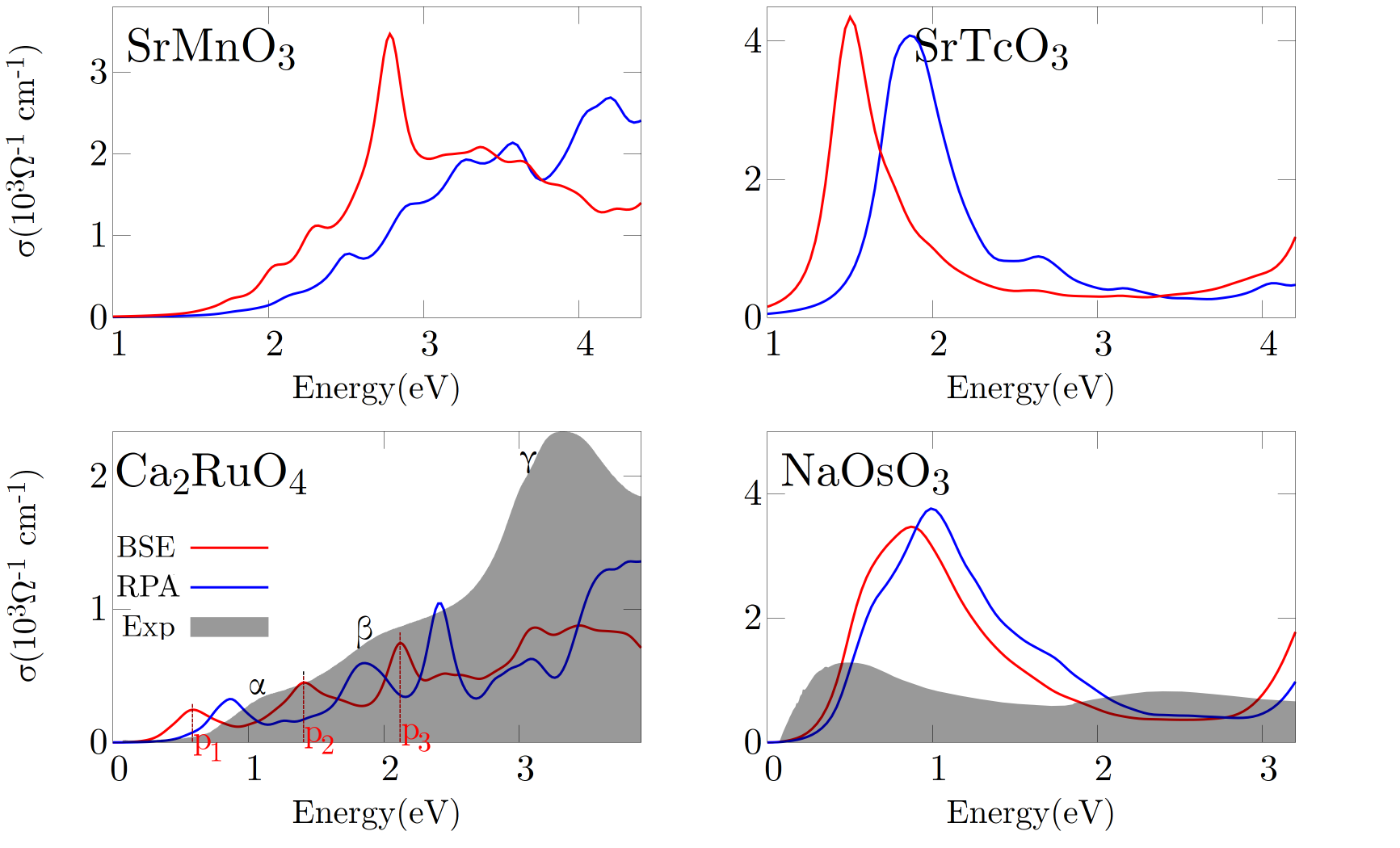}
\caption{\label{fig:RemainingPerov:OptCond}$\sigma(\omega)$ within the RPA ({\color{blue}blue line}), BSE approach ({\color{red}red line}) and experimental data; we could not find any experimental curve in literature for SrMnO$_3$ and SrTcO$_3$.
$\alpha$, $\beta$ and $\gamma$ label Ca$_2$RuO$_4$ experimental curve's peaks, while $\color{red} p_1$, $\color{red} p_2$ and $\color{red} p_3$ label the peaks of the BSE curve.
SrMnO$_3$ and Ca$_2$RuO$_4$ curves are obtained through a $4\times4\times2 | 2\times2\times2$ \textit{k}-averaging, SrTcO$_3$ through a $5\times5\times3 | 2\times2\times2$ one and NaOsO$_3$ through $3\times3\times3 | 2\times2\times2$.}
\end{figure*}

\subsection{\label{sec:RemainingPerov}C\lowercase{a}$_2$R\lowercase{u}O$_4$, N\lowercase{a}O\lowercase{s}O$_3$ and S\lowercase{r}TMO$_3$ (TM=M\lowercase{n}, T\lowercase{c})}

\subsubsection{\label{sec:Ca2RuO4:Comparison}Comparison between BSE and RPA spectra}

We complete the discussion of the results by reporting the analysis of the optical transitions for the remaining compounds: C\lowercase{a}$_2$R\lowercase{u}O$_4$, N\lowercase{a}O\lowercase{s}O$_3$, S\lowercase{r}MnO$_3$, and S\lowercase{r}T\lowercase{c}O$_3$ (the computed $\sigma(\omega)$ are collected in Fig.~\ref{fig:RemainingPerov:OptCond}).
For these compounds, a comparison with the measured optical conductivity is limited to NaOsO$_3$~\cite{NOO.Vecchio2013} and Ca$_2$RuO$_4$~\cite{CRO.Jung.PRL.91.056403} (to the best of our knowledge, we are not aware of any experimental spectra for SrMnO$_3$ and SrTcO$_3$). For those materials only a qualitative agreement between theory and experiment is achieved.

The experimental spectrum of Ca$_2$RuO$_4$  exhibits three distinct peaks: 2 weak shoulders labeled $\alpha$ and $\beta$ (following the nomenclature of Jung \textit{et al.}~\cite{CRO.Jung.PRL.91.056403}) and a third intense one at $\sim 3$ eV  designated as $\gamma$. 
These peaks are correctly identified by both RPA and BSE approaches, despite the lower intensities. $\sigma^{RPA}(\omega)$ underestimates the experimental onset, and the BSE slightly aggravates this discrepancy with a redshift of $\sim 0.2$ eV. The $\alpha$ and $\beta$ peaks predicted by BSE exhibit a slightly more pronounced redshift (respectively $\sim 0.3$ eV and $\sim 0.4$ eV). 

Despite retaining a transition metal of the same group, SrMnO$_3$ and SrTcO$_3$ exhibit rather different spectra. The 3$d$ SrMnO$_3$ perovskite displays a wide and multi-peaked structure between $2$ eV and $4$ eV. The excitonic corrections are prominent, with a significant enhancement of the peak at $2.8$ eV associated with a redshift of $\sim 0.6$ eV (evaluated at $\sigma \sim 2\times10^3\Omega^{-1}$cm$^{-1}$). For the 4$d$ perovskite SrTcO$_3$ BSE does not substantially modify the peaks intensity, but leads to a sizable redshift of about $0.4$ eV for the first peak (evaluated at $\sigma \sim 2\times10^3\Omega^{-1}$cm$^{-1}$).
The 5$d$ compound NaOsO$_3$ exhibits the highest $\epsilon_\infty$ within the dataset (see Fig.~\ref{fig:ExbTrend} suggesting a strong electronic screening) and the lowest excitonic redshift among all considered systems ($\sim 0.1$ eV, see Fig.~\ref{fig:ExbTrend} ). This is reminiscent of the BSE prediction for other 5$d$ systems (e.g. iridates~\cite{Liu.PRM.2.075003, app11062527}), indicating relatively weak excitonic effects in extended 5$d$ orbitals.

The calculated exciton binding energies, along with the parameters used for the constructing the model screening functions, are listed in Table~\ref{tab:RemainingPerv:Exb}. 
For this subset of materials, mBSE reproduces rather well the BSE binding energies $E_{xb}$. 

\begin{table}[h]
    \begin{tabularx}{\columnwidth}{@{}lcXXXX@{}}
    \toprule\toprule
                                                                          & k-mesh            & SrMnO$_3$ & SrTcO$_3$ & Ca$_2$RuO$_4$ & NaOsO$_3$ \\ \midrule
    \multirow{3}{*}{BSE}                                                  & $4\times4\times2$ &           &           & 0.199         &              \\
                                                                          & $5\times5\times3$ & 0.081     & 0.104     &               & 0.059        \\
    \multirow{3}{*}{\begin{tabular}[c]{@{}l@{}}mBSE\end{tabular}}         & $4\times4\times2$ &           &           & 0.165         &              \\
                                                                          & $5\times5\times3$ & 0.077     & 0.106     &               & 0.051        \\
    \% error                                                              &                   & 5         & 2         & 17            & 14             \\  \arrayrulecolor{black!45}\cmidrule{2-6}
    \multirow{2}{*}{\begin{tabular}[c]{@{}l@{}}mBSE\end{tabular}}         & $8\times8\times4$ & 0.034     &           & 0.118         &              \\ 
                                                                          & $9\times9\times6$ &           & 0.057     &               & 0.024        \\ \arrayrulecolor{black!45}\midrule
    $\epsilon_\infty^{-1}$                                                &                   & 0.088     & 0.071     & 0.087         & 0.031        \\
    $\lambda$                                                             &                   & 1.340     & 1.329     & 1.225         & 1.109        \\ \arrayrulecolor{black}\bottomrule\bottomrule
    \end{tabularx}
    \caption{\label{tab:RemainingPerv:Exb} Exciton binding energies $E_{xb}$ in eV for Ca$_2$RuO$_4$, NaOsO$_3$ and STMO$_3$ (TM=Mn,Tc), calculated through the mBSE and  BSE approaches. 
    Conventions used are the same as in Table~\ref{tab:CubicPerov:Exb}. 
    }
\end{table}

\begin{figure*}[t!]
\centering
\includegraphics[width=\textwidth]{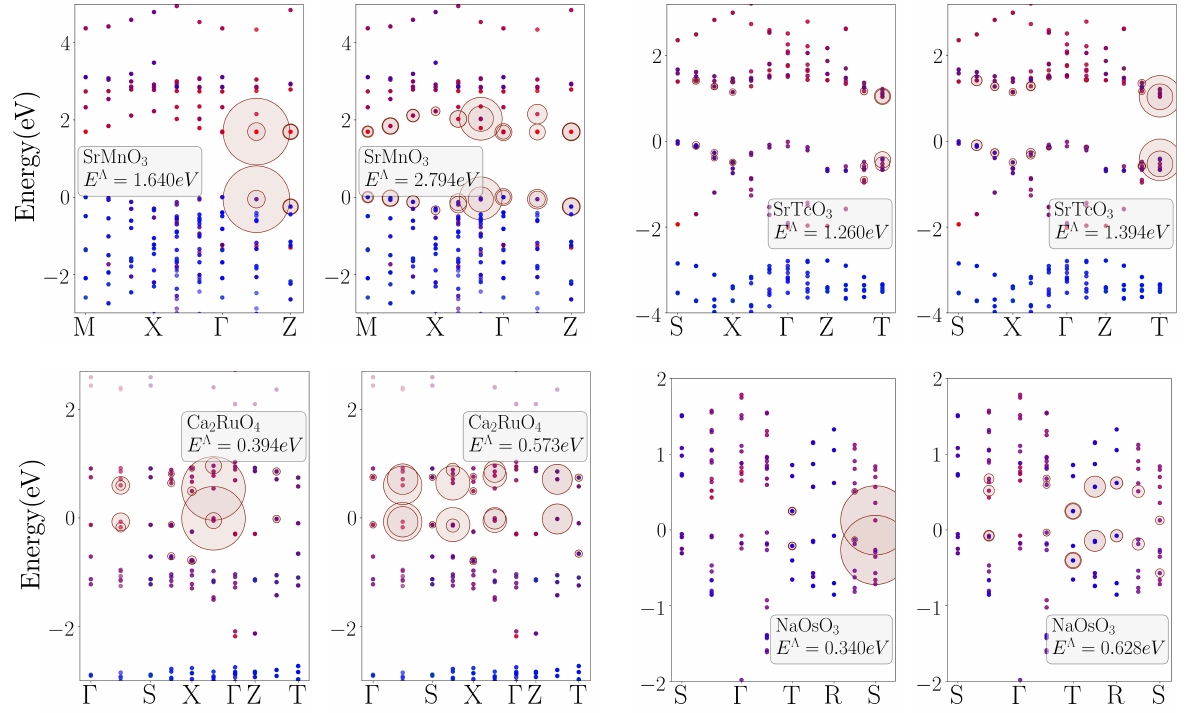}
	\caption{\label{fig:RemainingPerov:Fatbanb}
	Fat band pictures for SrMnO$_3$, SrTcO$_3$, Ca$_2$RuO$_4$ and NaOsO$_3$.
	The left column refers to the \textit{first} non-dark eigenvector of each materials; the right column to the eigenvectors associated with the highest peaks in the first structures.
	Color codings and labellings are analogous to those adopted in Fig.~\ref{fig:LaSeries:Fatbanb}: {\color{blue}blue for O$-p$}, {\color{red}red for TM$-t_{2g}$} and {\color{green}green for Sr/Ca/Na$-d$}.}
\end{figure*}

\subsubsection{\label{sec:RemainingPerov:MainOptTrans}Origin and character of main optical transitions}

We discuss the nature of the main transitions based on the fat bands analysis shown in Fig.~\ref{fig:RemainingPerov:Fatbanb}. Similar to the previous cases, we focus our analysis on the first non-dark excitations and on the main peaks in the first part of the optical spectra.

For SrTcO$_3$ both the optical gap and the sharp peak at $1.4$ eV exhibit a clear Mott character. Although the greater contributions to the excitonic wavefunctions originate from the $Z-T$ direction, the excitonic wavefunctions themselves are delocalized in the BZ and their amplitudes are suppressed at the $\Gamma$ point, as expected from Mott-type $d-d$ transitions (see discussion for the La series).
\newline As already mentioned, SrMnO$_3$ displays marked differences: the uppermost valence bands exhibit a strong admixture of O$-p$ and Mn$-d$ (with a O$-p$ percentage varying between $18\%$ along the $\Gamma-X$ direction and $46\%$ along $M-R$)  indicating an intermediate CT/MH nature of the optical gap~\cite{Ergorenc.PRM.2.024601}. The associated  wavefunction is more localized than the one calculated for SrTcO$_3$, with strong contributions only around the $\Gamma-X$ direction. 

The low-energy electronic structure of Ca$_2$RuO$_4$ has been widely studied and discussed~\cite{CRO.Sutter.Nature.2017,CRO.Jung.PRL.91.056403,CRO.Mizokawa.PRL.87.077202,CRO.Fatuzzo.PRB.91.155104,CRO.Gorelov.PRL.104.226401,CRO.PRB.95.075145,CRO.PRB.101.205128,CRO.PRB.88.104428,CRO.PRL.88.017201,CRO.PRB.95.075145}. 
The 4$d$ electrons occupy the $t_{2g}$ orbitals, with the $e_g$ states well separated in energy and completely empty.
Our $G_0W_0$ data, in agreement with previous computational studies~\cite{CRO.Jung.PRL.91.056403,CRO.Fang.PRB.69.045116}, describes a $t_{2g}$ manifold subjected to an further splitting, with the $d_{xy}$ dominantly occupied and the $d_{xz}$/$d_{yz}$  mixed and partially filled.
The $t_{2g}$ splitting has been associated with different mechanisms, including the rotation and tilting of the RuO$_6$ octahedra~\cite{CRO.Mizokawa.PRL.87.077202}, spin-orbit interaction~\cite{CRO.Fatuzzo.PRB.91.155104}, c-axis contraction and crystal field stabilization~\cite{CRO.Gorelov.PRL.104.226401,CRO.PRB.95.075145}. 
The O$-p$ percentage varies between $14\%$ and $25\%$ for the highest valence band and between $20\%$ and $24\%$ for the lowest conduction band, indicating an admixture of $p$ and $d$ states with a predominantly MH $d-d$ optical gap~\cite{Ergorenc.PRM.2.024601}.
Our data suggest that the $\alpha$ peak is determined by transitions between filled and empty $t_{2g}$ bands, more specifically between filled $d_{xy}$ states and conduction $d_{xz}$/$d_{yz}$ states.
The $\beta$ peak at $\sim 1.5$ eV can be assigned to transitions between the splitted $d_{xz}$/$d_{yz}$ manifolds, whereas the $\gamma$ peak at 2.2 eV is established by excitations from $d_{xy}$ orbitals just below the Fermi energy to Ru-$e_g$ states $> 2$ eV above the Fermi energy (see Fig.~\ref{fig:CRO:FatbandPictures}).
The wide structure around $3$ eV in the BSE spectrum is determined by excitations from the O$-p$ bands near $-3$ eV under the Fermi energy towards the conduction $d_{xz}$/$d_{yz}$ orbitals.

\begin{figure}[!ht]
	\includegraphics[width=\columnwidth]{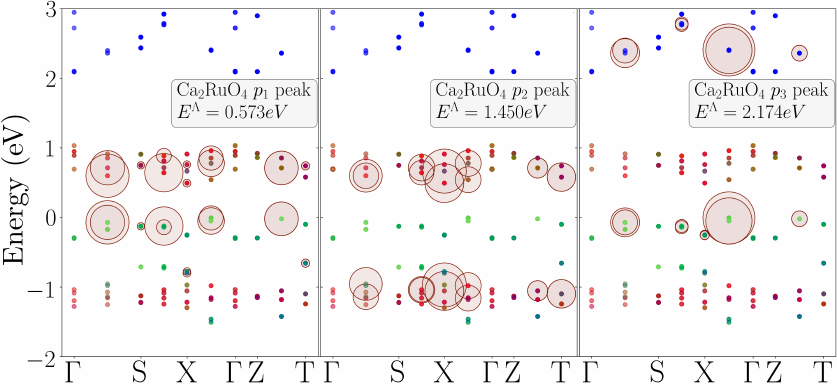} 
	\caption{\label{fig:CRO:FatbandPictures}Fat band picture for $A_{ \mathbf{k} v c}$ related to the first three BSE peaks of Ca$_2$RuO$_4$. {\color{green}Green color for Ru$-d_{xy}$}, {\color{red}red for mixed Ru$-d_{xz}/d_{yz}$}, {\color{blue}blue for Ru$-e_g$}.}
\end{figure}

\section{Summary and Conclusions}\label{sec:conclude}
We have performed a systematic investigation of the optical properties of a selected set of transition metal oxide perovskites by \textit{ab initio } $G_0W_0$+BSE.  The fourteen compounds were selected to constitute a minimal dataset, representative of the variety of structural and electronic  properties characteristic of this class of perovskites.
\newline The solution of the BSE equation proves to be decisive to obtain a quantitative agreement between the theoretical and experimental spectra for the cubic perovskites SrTiO$_3$, SrHfO$_3$, SrZrO$_3$ and KTaO$_3$. A pronounced spectral weight transfer is visible in their optical conductivity lineshapes (with an average redshift at the onset of  $1.03$ eV) and can be related to prominent excitonic contributions. This confirms and extends the previous studies on SrTiO$_3$.
\newline To investigate the origin of the main structures of the spectra we analyzed the e-h coupling coefficients associated with the most intense oscillator strengths.
The contributions from transitions towards different conduction band manifolds are examined, and the role of band overlapping in SrZrO$_3$ and SrHfO$_3$ is discussed.

Comparisons with the reference experimental data have been discussed for the La series, NaOsO$_3$ and Ca$_2$RuO$_4$. 
The main structures visible in the experimental spectra are correctly identified in both RPA and BSE approaches with similar lineshapes. The excitonic corrections for this subset can be summarized as a redshift of the entire spectra, with an average  value of $0.5$ eV and reduced peak enhancements (with LaFeO$_3$ as partial exception).
The BSE approach, however, consistently underestimates the experimental onset by $0.3$ eV - $1.0$ eV and incorrectly describes the mixed MH/CT nature of LaCrO$_3$'s optical gap.
Additional calculations based on hybrid functionals displayed significant improvements with respect to the overall optical conductivity lineshape, optical bandgap value and character. Hence employing hybrid functionals as a starting point may represent a promising route for further investigations.
\newline The exciton binding energies were calculated through a mBSE approach. This method introduces two approximations: a scissor operator to mimic the quasi-particle shifts and model dielectric function to determine the dielectric screening. The lower computational cost allows to achieve converged $E_{xb}$ with respect to the k-point mesh. To assess the validity of this approach, we benchmarked the mBSE calculated $E_{xb}$ against the $G_0W_0$+BSE values obtained with the same k-point mesh. The BSE values are very well reproduced for the cubic subset (with a mean absolute percentage error (MAPE) of $6\%$)  and for SrMnO$_3$, SrTcO$_3$, Ca$_2$RuO$_4$ and NaOsO$_3$ (with a MAPE of $9\%$). The discrepancies for the La series are larger, with a MAPE of $20\%$; the largest outlier is LaVO$_3$, primarily due to the diagonal dielectric screening approximation.

The overall agreement with experimental data is satisfactory, also considering the technical difficulties that hamper a precise measurement of optical spectra for TMO perovskites, and the tendency of this class of materials to be subjected to chemical defect (e.g. oxygen vacancies or presence of TM impurities). Our reported data represents a first comprehensive map of the optical and excitonic properties of complex TMO and should serve as a reference for future calculations and experiments.

\begin{acknowledgments}
The computational results presented have been achieved using the Vienna Scientific Cluster (VSC) and the CINECA HPC infrastructure. 
\end{acknowledgments}

\clearpage
\bibliography{ms}

\end{document}


\title{Supplementary Material for ``Optical properties of transition metal oxide perovskites by the Bethe-Salpeter equation''}
\author{Lorenzo Varrassi, Peitao Liu, Zeynep Erg{\"o}renc, Menno Bokdam, Georg Kresse, Cesare Franchini}
\maketitle
\appendix\section{Details on the reference optical experiments}
In this section the experimental setups employed for the various optical spectra measurements are listed.
\begin{table}[h]
\begin{tabularx}{\textwidth}{@{}Xl@{}}
\toprule
Perovskite    & Description                                                                  \\ \midrule
SrTiO$_3$~\cite{SZO.OptCond}     & Near normal incident reflectivity spectra measured at room temperature.      \\
SrZrO$_3$~\cite{SZO.OptCond} &
  \begin{tabular}[c]{@{}l@{}}Near normal incident reflectivity spectra measured at room temperature \\ in the energy region 5meV-30eV.\end{tabular} \\
SrHfO$_3$\cite{SHO.OptCond} &
  \begin{tabular}[c]{@{}l@{}}Near normal incident reflectivity spectra measured for 5 meV-30 eV \\ at room temperature. Conventional Fourier Transform spectrophotometer \\ used between 5meV and 0.8eV, grating spectrometers used above 0.6 eV.\end{tabular} \\
KTaO$_3$~\cite{KTO.OptCond}      & Near normal incident reflectivity spectra measured at room temperature.      \\
\begin{tabular}[t]{@{}l@{}}La series~\cite{Arima.PRB.48.17006}\\ (except \\ LaMnO$_3$)\end{tabular} &
  \begin{tabular}[t]{@{}l@{}}Near normal incident reflectivity spectra. Fourier Trasform interferometer \\ used for 0.015 - 0.8 eV, grating type monochromator used for 0.6 - 36 eV.\end{tabular} \\
LaMnO$_3$~\cite{LMO.OptCond} &
  \begin{tabular}[t]{@{}l@{}}Near normal incidence reflectivity spectra measured at room temperature. \\ A Fourier Transform spectrophotometer used between 5 meV and 0.8 eV, \\ grating monochromator used between 0.6 and 7.0 eV.\end{tabular} \\
\begin{tabular}[t]{@{}l@{}}Additional\\reference \\ for LaVO$_3$~\cite{Miyasaka.JPSJ.71.2086}\end{tabular} &\begin{tabular}[t]{@{}l@{}}Reflectivity measurements at 10K. Fourier-transform interferometer for \\  the  0.06 eV - 0.8 eV range, grating spectrometers for the 0.6 eV - 4.0 eV region.\end{tabular} \\
SrMnO$_3$     & No experimental data available.                                              \\
SrTcO$_3$     & No experimental data available.                                              \\
Ca$_2$RuO$_4$~\cite{CRO.OptCond} & Near normal incident reflectivity spectra measured for 0.06 - 36 eV at 10 K. \\
NaOsO$_3$~\cite{NOO.OptCond} &
  \begin{tabular}[c]{@{}l@{}}Near normal reflectance measurement using a Michelson interferometer \\ at 5 K in the frequency range 10 $cm^{-1}$ to 15000 $cm^{-1}$.\end{tabular} \\ \bottomrule
\end{tabularx}
\caption{literature references and experimental setups for the optical spectra data cited in the paper.}
\end{table}

\newpage\appendix\section{Crystal structures}
In this section the atomic positions (in the POSCAR format used in VASP) of all materials studied are listed. These positions are obtained from experimental studies: SrTiO$_3$ ~\cite{STO.CrystalPaper}, SrZrO$_3$ ~\cite{SZO.Smith:a02873}, SrHfO$_3$ ~\cite{SHO.PhysRevB.60.2972}, KTaO$_3$ ~\cite{KTO.PhysRevLett.88.097601},LaScO$_3$ ~\cite{LSCO.Geller:a01971}, LaTiO$_3$~\cite{LTO.PhysRevB.68.060401}, LaVO$_3$ ~\cite{LVO.BORDET1993253}, LaCrO$_3$ ~\cite{LCO.KHATTAK1977463}, LaMnO$_3$ ~\cite{LMO.ELEMANS1971238},  LaFeO$_3$ ~\cite{LFO.DANN1994134},SrMnO$_3$ ~\cite{SMO.PhysRevB.74.144102}, SrTcO$_3$ ~\cite{STcO.PhysRevLett.106.067201}, Ca$_2$RuO$_4$ ~\cite{CRO.PhysRevB.58.847}, NaOsO$_3$ ~\cite{NOO.PhysRevB.80.161104}. \newline For SrMnO$_3$ we employed the calculated lattice constant for the G-AFM phase $3.824 \si{\angstrom}$ ~\cite{SMO.PhysRevB.74.144102} instead of the corresponding experimental value ($3.800 \si{\angstrom}$) ~\cite{SMOExp}.

\begin{tasks}(2)
\task[]\texttt{KTaO3\newline
3.988\newline
 1.000000000  0.000000000  0.000000000\newline
 0.000000000  1.000000000  0.000000000\newline
 0.000000000  0.000000000  1.000000000\newline
K  Ta O\newline
1  1  3\newline
Direct\newline
 0.000000000  0.000000000  0.000000000\newline
 0.500000000  0.500000000  0.500000000\newline
 0.500000000  0.500000000  0.000000000\newline
 0.500000000  0.000000000  0.500000000\newline
 0.000000000  0.500000000  0.500000000}
\task[]\texttt{SrHfO3\newline
4.1138\newline 
 1.000000000  0.000000000  0.000000000\newline
 0.000000000  1.000000000  0.000000000\newline
 0.000000000  0.000000000  1.000000000\newline
Sr Hf O\newline
1  1  3\newline
Direct\newline
 0.000000000  0.000000000  0.000000000\newline
 0.500000000  0.500000000  0.500000000\newline
 0.500000000  0.500000000  0.000000000\newline
 0.500000000  0.000000000  0.500000000\newline
 0.000000000  0.500000000  0.500000000}
\task[]\texttt{SrTiO3\newline
 3.905\newline
  1.000000000  0.000000000  0.00000000\newline
  0.000000000  1.000000000  0.00000000\newline
  0.000000000  0.000000000  1.00000000\newline
Sr Ti O\newline
1 1 3\newline
Direct\newline
 0.500000000  0.500000000  0.500000000\newline
 0.000000000  0.000000000  0.000000000\newline
 0.500000000  0.000000000  0.000000000\newline
 0.000000000  0.000000000  0.500000000\newline
 0.000000000  0.500000000  0.000000000}
\task[]\texttt{SrZrO3\newline
4.109\newline
 1.000000000  0.000000000  0.000000000\newline
 0.000000000  1.000000000  0.000000000\newline
 0.000000000  0.000000000  1.000000000\newline
Sr Zr O\newline
1  1  3\newline
Direct\newline
 0.000000000  0.000000000  0.000000000\newline
 0.500000000  0.500000000  0.500000000\newline
 0.500000000  0.500000000  0.000000000\newline
 0.500000000  0.000000000  0.500000000\newline
 0.000000000  0.500000000  0.500000000}
\task[]\texttt{LaScO3\newline
1.00\newline
 5.680300236  0.000000000  0.000000000\newline
 0.000000000  5.790699959  0.000000000\newline
 0.000000000  0.000000000  8.094499588\newline
La   Sc   O\newline
4    4   12\newline
Direct\newline
 0.957199991  0.010000000  0.750000000\newline
 0.042800009  0.990000010  0.250000000\newline
 0.457199991  0.490000010  0.250000000\newline
 0.542800009  0.509999990  0.750000000\newline
 0.500000000  0.000000000  0.000000000\newline
 0.000000000  0.500000000  0.000000000\newline
 0.500000000  0.000000000  0.500000000\newline
 0.000000000  0.500000000  0.500000000\newline
 0.527700007  0.903199971  0.750000000\newline
 0.472299993  0.096800029  0.250000000\newline
 0.027700007  0.596800029  0.250000000\newline
 0.972299993  0.403199971  0.750000000\newline
 0.295800000  0.707300007  0.052100003\newline
 0.704200029  0.292699993  0.947899997\newline
 0.795799971  0.792699993  0.947899997\newline
 0.204200000  0.207300007  0.052100003\newline
 0.704200029  0.292699993  0.552100003\newline
 0.295800000  0.707300007  0.447899997\newline
 0.204200000  0.207300007  0.447899997\newline
 0.795799971  0.792699993  0.552100003}
\task[]\texttt{LaTiO3\newline
1.00\newline
 5.588500023  0.000000000  0.000000000 \newline
 0.000000000  0.000000000  5.643499851\newline
 0.000000000  7.900599957  0.000000000\newline 
La  Ti  O\newline
4   4   12\newline
Direct\newline
 0.049100000  0.493000001  0.250000000\newline 
 0.950900018  0.506999969  0.750000000\newline 
 0.450899988  0.993000031  0.750000000\newline 
 0.549099982  0.006999999  0.250000000\newline
 0.000000000  0.000000000  0.000000000\newline 
 0.500000000  0.500000000  0.000000000\newline
 0.000000000  0.000000000  0.500000000\newline 
 0.500000000  0.500000000  0.500000000\newline 
 0.493999988  0.581300020  0.250000000\newline 
 0.506000042  0.418699980  0.750000000\newline
 0.006000012  0.081300020  0.750000000\newline 
 0.993999958  0.918699980  0.250000000\newline 
 0.294299990  0.209199995  0.042800002\newline
 0.705700040  0.790799975  0.957199991\newline
 0.205700010  0.709200025  0.957199991\newline
 0.794299960  0.290800005  0.042800002\newline 
 0.705700040  0.790799975  0.542800009\newline 
 0.294299990  0.209199995  0.457199991\newline 
 0.794299960  0.290800005  0.457199991\newline 
 0.205700010  0.709200025  0.542800009}
\task[]\texttt{LaVO3\newline
1.00\newline
 0.000000000  0.000000000  5.562300205\newline
 5.591700554  0.000000000  0.000000000\newline
-0.017452219  7.751580142  0.000000000\newline
La    V    O\newline
4    4   12\newline
Direct\newline
 0.991699994  0.034100000  0.250000000\newline
 0.008300006  0.965900004  0.750000000\newline
 0.491699994  0.465900004  0.750000000\newline 
 0.508300006  0.534099996  0.250000000\newline 
 0.000000000  0.500000000  0.000000000\newline 
 0.500000000  0.000000000  0.000000000\newline
 0.500000000  0.000000000  0.500000000\newline 
 0.000000000  0.500000000  0.500000000\newline 
 0.007000000  0.490000010  0.250000000\newline 
 0.992999971  0.509999990  0.750000000\newline 
 0.507000029  0.009999990  0.750000000\newline 
 0.493000001  0.990000010  0.250000000\newline 
 0.714399993  0.277599990  0.038100000\newline 
 0.285600007  0.722400010  0.961899996\newline 
 0.214399993  0.222400010  0.961899996\newline 
 0.785600007  0.777599990  0.038100000\newline 
 0.283299983  0.710399985  0.538900018\newline 
 0.716700017  0.289600015  0.461099982\newline 
 0.783299983  0.789600015  0.461099982\newline
 0.216700017  0.210399985  0.538900018}       
\task[]\texttt{LaCrO3\newline 
1.00\newline
 5.480299950  0.000000000  0.000000000\newline
 0.000000000  0.000000000  5.516799927\newline
 0.000000000  7.759900093  0.000000000\newline
La   Cr   O\newline 
4     4    12\newline
Direct\newline
 0.027005915  0.995296785  0.250000000\newline 
 0.972994064  0.004703207  0.750000000\newline 
 0.472994094  0.495296785  0.750000000\newline 
 0.527005935  0.504703215  0.250000000\newline
 0.000000000  0.500000000  0.000000000\newline 
 0.500000000  0.000000000  0.000000000\newline 
 0.000000000  0.500000000  0.500000000\newline 
 0.500000000  0.000000000  0.500000000\newline
 0.490499144  0.068467118  0.250000000\newline 
 0.509500885  0.931532860  0.750000000\newline 
 0.009500855  0.568467140  0.750000000\newline 
 0.990499114  0.431532889  0.250000000\newline 
 0.219218757  0.219511552  0.536131336\newline 
 0.780781212  0.780488448  0.463868664\newline 
 0.280781242  0.719511552  0.463868664\newline 
 0.719218787  0.280488448  0.536131336\newline 
 0.780781212  0.780488448  0.036131336\newline 
 0.219218757  0.219511552  0.963868664\newline 
 0.719218787  0.280488448  0.963868664\newline 
 0.280781242  0.719511552  0.036131336}
\task[]\texttt{LaMnO3\newline
1.00\newline
 0.000000000  0.000000000  5.532000000\newline
 5.742000000  0.000000000  0.000000000\newline
 0.000000000  7.668000000  0.000000000\newline
La Mn O\newline
4  4 12\newline
Direct\newline
 0.010000000  0.549000025  0.250000000\newline 
 0.510000000 -0.049000025 -0.250000000\newline 
-0.010000000 -0.549000025  0.750000000\newline 
 0.490000000  1.049000025  0.250000000\newline
 0.000000000  0.000000000  0.000000000\newline 
 0.500000000  0.500000000  0.000000000\newline 
 0.000000000  0.000000000  0.500000000\newline 
 0.500000000  0.500000000  0.500000000\newline 
 0.224000007  0.308999985  0.039000001\newline 
 0.724000007  0.191000015 -0.039000001\newline 
-0.224000007 -0.308999985  0.539000001\newline 
 0.275999993  0.808999985  0.460999999\newline 
-0.224000007 -0.308999985 -0.039000001\newline
 0.275999993  0.808999985  0.039000001\newline 
 0.224000007  0.308999985  0.460999999\newline 
 0.724000007  0.191000015  0.539000001\newline 
-0.070000000 -0.014000000  0.250000000\newline 
 0.430000000  0.514000000 -0.250000000\newline 
 0.070000000  0.014000000  0.750000000\newline 
 0.570000000  0.486000000  0.250000000}
\task[]\texttt{LaFeO3\newline
1.0\newline
 0.000000000  0.000000000  5.556600094\newline
 5.568200111  0.000000000  0.000000000\newline
 0.000000000  7.856900215  0.000000000\newline
La   Fe  O\newline
4    4   12\newline
Direct\newline
 0.993499994  0.029600024  0.250000000\newline 
 0.006500006  0.970399976  0.750000000\newline
 0.493499994  0.470399976  0.750000000\newline 
 0.506500006  0.529600024  0.250000000\newline 
 0.500000000  0.000000000  0.000000000\newline
 0.000000000  0.500000000  0.000000000\newline 
 0.500000000  0.000000000  0.500000000\newline 
 0.000000000  0.500000000  0.500000000\newline 
 0.071399987  0.486400008  0.250000000\newline 
 0.928600013  0.513599992  0.750000000\newline
 0.571399987  0.013599992  0.750000000\newline 
 0.428600013  0.986400008  0.250000000\newline
 0.286000013  0.716099977  0.961099982\newline 
 0.213999987  0.216099977  0.961099982\newline
 0.786000013  0.783900023  0.038900018\newline 
 0.286000013  0.716099977  0.538900018\newline 
 0.713999987  0.283900023  0.461099982\newline 
 0.786000013  0.783900023  0.461099982\newline 
 0.213999987  0.216099977  0.538900018}  
\task[]\texttt{SrMnO3\newline
1.00\newline
 5.407000065  0.000000000  0.000000000\newline
 0.000000000  5.407000065  0.000000000\newline
 0.000000000  0.000000000  7.647999764\newline
Sr   Mn  O\newline
4    4   12\newline
Direct\newline
 0.000000000  0.000000000  0.000000000\newline    
 0.500000000  0.500000000  0.000000000\newline   
 0.500000000  0.500000000  0.500000000\newline 
 0.000000000  0.000000000  0.500000000\newline  
 0.500000000  0.000000000  0.250000000\newline
 0.000000000  0.500000000  0.750000000\newline  
 0.000000000  0.500000000  0.250000000\newline  
 0.500000000  0.000000000  0.750000000\newline 
 0.500000000  0.000000000  0.000000000\newline   
 0.000000000  0.500000000  0.000000000\newline
 0.500000000  0.000000000  0.500000000\newline   
 0.000000000  0.500000000  0.500000000\newline   
 0.250000000  0.250000000  0.250000000\newline   
 0.250000000  0.750000000  0.250000000\newline
 0.750000000  0.250000000  0.250000000\newline   
 0.750000000  0.750000000  0.250000000\newline   
 0.750000000  0.750000000  0.750000000\newline   
 0.750000000  0.250000000  0.750000000\newline   
 0.250000000  0.750000000  0.750000000\newline   
 0.250000000  0.250000000  0.750000000}
\task[]\texttt{SrTcO3\newline
1.00\newline
 5.535472762  0.000000000  0.000000000\newline
 0.000000000  0.000000000  5.587831295\newline
 0.000000000  7.845031008  0.000000000\newline
Sr   Tc   O\newline
4     4   12\newline
Direct\newline
 0.001923854  0.498281285  0.250000300\newline
 0.998075582  0.501720142  0.749999629\newline
 0.498076766  0.998279674  0.749999629\newline
 0.501924474  0.001720326  0.250000300\newline
 0.000000000  0.000000000  0.000000000\newline
 0.500000621  0.499999817  0.000000000\newline
 0.000000000  0.000000000  0.500000600\newline
 0.500000621  0.499999817  0.500000600\newline
 0.253225037  0.247037929  0.025306312\newline
 0.746774399  0.752961705  0.974693617\newline
 0.246775584  0.747037745  0.974693617\newline
 0.753225657  0.252961888  0.025306312\newline
 0.746774399  0.752961705  0.525305640\newline
 0.253225037  0.247037929  0.474694289\newline
 0.753225657  0.252961888  0.474694289\newline
 0.246775584  0.747037745  0.525305640\newline
 0.499321610  0.549508287  0.250000300\newline
 0.500677826  0.450491346  0.749999629\newline
 0.000677205  0.049508470  0.749999629\newline
 0.999322233  0.950491163  0.250000300}
\end{tasks}
\begin{tasks}(2)
\task[] \texttt{Ca2RuO4\newline
1.00\newline
 5.387700081  0.000000000   0.000000000\newline
 0.000000000  5.632299900   0.000000000\newline
 0.000000000  0.000000000  11.746299744\newline
Ca   Ru   O\newline 
8    4    16\newline
Direct\newline
 0.005436689  0.062439183  0.351941048\newline
 0.994563313  0.937560814  0.648058922\newline
 0.494563313  0.937560814  0.851941078\newline
 0.505436686  0.062439183  0.148058952\newline
 0.994563313  0.562439186  0.148058952\newline
 0.005436688  0.437560814  0.851941078\newline
 0.505436686  0.437560814  0.648058922\newline
 0.494563313  0.562439186  0.351941048\newline
 0.000000000  0.000000000  0.000000000\newline
 0.500000000  0.000000000  0.500000000\newline
 0.000000000  0.500000000  0.500000000\newline
 0.500000000  0.500000000  0.000000000\newline
 0.191394478  0.303978940  0.028666039\newline
 0.808605507  0.696021060  0.971333962\newline
 0.308605507  0.696021060  0.528666038\newline
 0.691394492  0.303978940  0.471333962\newline
 0.808605508  0.803978940  0.471333962\newline
 0.191394478  0.196021060  0.528666038\newline
 0.691394492  0.196021060  0.971333962\newline
 0.308605508  0.803978940  0.028666039\newline
 0.926452660  0.975749063  0.165536388\newline
 0.073547340  0.024250937  0.834463597\newline
 0.573547340  0.024250937  0.665536403}

\task[]\texttt{0.426452660  0.975749063  0.334463597\newline
 0.073547340  0.475749063  0.334463597\newline
 0.926452660  0.524250937  0.665536403\newline
 0.426452660  0.524250936  0.834463597\newline
 0.573547340  0.475749063  0.165536388}

\task[]\texttt{NaOsO3\newline
 7.58038000\newline
 0.00000000    0.00000000    0.70288956\newline
 0.71028101    0.00000000    0.00000000\newline
 0.00000000    1.00000000    0.00000000\newline
Na O  Os\newline
4  12  4\newline
Direct\newline
 0.99350000  0.03280000  0.25000000\newline
 0.49350000  0.46720000  0.75000000\newline
 0.00650000  0.96720000  0.75000000\newline
 0.50650000  0.53280000  0.25000000\newline
 0.71120000  0.28810000  0.03940000\newline   
 0.21120000  0.21190000  0.96060000\newline   
 0.28880000  0.71190000  0.53940000\newline   
 0.78880000  0.78810000  0.46060000\newline   
 0.28880000  0.71190000  0.96060000\newline   
 0.78880000  0.78810000  0.03940000\newline   
 0.71120000  0.28810000  0.46060000\newline   
 0.21120000  0.21190000  0.53940000\newline   
 0.08080000  0.48340000  0.25000000\newline   
 0.58080000  0.01660000  0.75000000\newline   
 0.91920000  0.51660000  0.75000000\newline   
 0.41920000  0.98340000  0.25000000\newline   
 0.50000000  0.00000000  0.00000000\newline  
 0.00000000  0.50000000  0.00000000\newline   
 0.50000000  0.00000000  0.50000000\newline   
 0.00000000  0.50000000  0.50000000}
\end{tasks}

\newpage\section{Scissor operators}
In this section the parameters and the k-point meshes used for the model-BSE calculations of the exciton binding energies $E_{xb}$ are collected:  
\begin{table}[h]
\begin{tabularx}{\columnwidth}{@{}lXXXX@{}}
\toprule\toprule
              & k-mesh                   & $\epsilon_\infty^{-1}$ & $\lambda$ & Scissor.op \\ \midrule
SrTiO$_3$     & $20 \times 20 \times 20$ & 0.1654                 & 1.4629    & 1.645      \\
SrZrO$_3$     & $20 \times 20 \times 20$ & 0.2309                 & 1.4573    & 0.9439     \\
SrHfO$_3$     & $20 \times 20 \times 20$ & 0.2419                 & 1.4478    & 1.9131     \\
KTaO$_3$      & $20 \times 20 \times 20$ & 0.1946                 & 1.4200    & 1.3946     \\
LaScO$_3$     & $10 \times 10 \times 6$  & 0.2013                 & 1.4621    & 1.7000     \\
LaTiO$_3$     & $10 \times 10 \times 6$  & 0.1202                 & 1.3491    & 0.3674     \\
LaVO$_3$      & $10 \times 10 \times 6$  & 0.1220                 & 1.4201    & 0.3275     \\
LaCrO$_3$     & $10 \times 10 \times 6$  & 0.1468                 & 1.3931    & 1.6057     \\
LaMnO$_3$     & $10 \times 10 \times 6$  & 0.1068                 & 1.3353    & 0.8497     \\
LaFeO$_3$     & $10 \times 10 \times 6$  & 0.1031                 & 1.3362    & 1.0644     \\
SrMnO$_3$     & $8\times 8\times 4$      & 0.0875                 & 1.3404    & 1.2873     \\
SrTcO$_3$     & $9\times 9\times 6$      & 0.0713                 & 1.3292    & 0.6018     \\
Ca$_2$RuO$_4$ & $8\times 8\times 4$      & 0.0870                 & 1.2247    & 0.4796     \\
NaOsO$_3$     & $9\times 9\times 6$      & 0.0308                 & 1.1092    & 0.2601     \\ \bottomrule\bottomrule
\end{tabularx}
\caption{\label{tab:appendix:mBSEParameters} Parameters for the model-BSE calculations of the converged exciton binding energies. 
The inverse static dielectric constants and screening lengths $\lambda$ ($\si{\angstrom}^{-1}$) used for the model dielectric function are given; Scissor.op stands for the scissor operator needed to approximate the $G_0W_0$ band structures.}
\end{table}
   
\newpage\section{\label{sectionC}K-averaging for SrTiO$_3$}
\begin{figure}[!hbt]
	\includegraphics[width=\columnwidth]{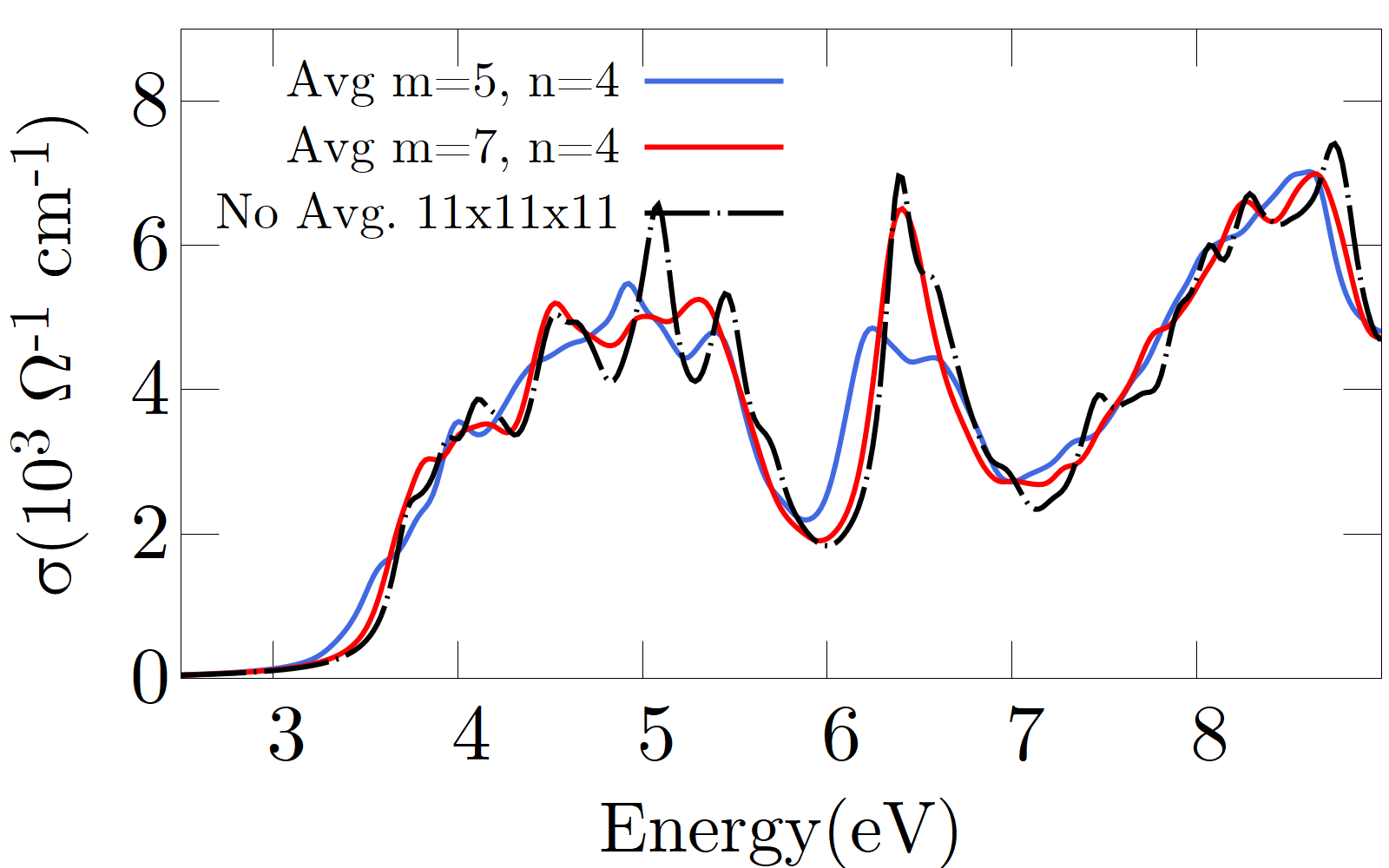} 
	\caption{\label{fig:appendix:STOAveraging} SrTiO$_3$ BSE spectra calculated for different k-averaging parameters $n$, $m$ and compared with a standard BSE calculation on an $11 \times 11 \times 11$ k-mesh.	}
\end{figure}

In Fig. \ref{fig:appendix:STOAveraging} we compare the spectra associated with two different k-averaging specifications to standard (single) BSE one to illustrate the presence of spurious peak artifacts (associated to the Coulomb kernel truncation approximation discussed in \cite{Sander.PRB.92.045209}). The $m=5$, $n=4$ exhibits a clear suppression of the $6.4$ eV sharp and narrow peak with respect to the standard BSE calculation, which is restored by the $m=7$, $n=4$ curve.

\bibliography{supplement}